\documentclass[11pt]{article}

\usepackage[tbtags]{amsmath}
\usepackage{amsthm}
\usepackage{amssymb}
\usepackage{amsfonts}
  \theoremstyle{plain}
  \newtheorem{theorem}{Theorem}
  \newtheorem{lemma}[theorem]{Lemma}

  \theoremstyle{definition}
  \newtheorem{definition}[theorem]{Definition}

  \theoremstyle{remark}
  \newtheorem*{remark}{Remark}
  \theoremstyle{plain}
  \newtheorem*{theorem*}{Theorem}
  \newtheorem*{lemma*}{Lemma}
  \newtheorem*{corollary*}{Corollary}
  \newtheorem*{proposition*}{Proposition}
  \newtheorem*{claim*}{Claim}

\newenvironment{step}
  {
    \begin{enumerate}

  }
  {\end{enumerate}}

\newenvironment{algorithm*}[1]
  {
    \begin{center}
      \hrulefill\\
      \textbf{#1}
  }
  {
    \vspace{-1\baselineskip}
    \hrulefill
    \end{center}
  }

\newenvironment{protocol*}[1]
  {
    \begin{center}
      \hrulefill\\
      \textbf{#1}
  }
  {
    \vspace{-1\baselineskip}
    \hrulefill
    \end{center}
  }

\newcommand{\calA}{\mathcal{A}}
\newcommand{\calB}{\mathcal{B}}

\newcommand{\calH}{\mathcal{H}}

\newcommand{\calK}{\mathcal{K}}

\newcommand{\calM}{\mathcal{M}}
\newcommand{\calN}{\mathcal{N}}

\newcommand{\calP}{\mathcal{P}}

\newcommand{\calS}{\mathcal{S}}

\newcommand{\calV}{\mathcal{V}}
\newcommand{\calW}{\mathcal{W}}
\newcommand{\calX}{\mathcal{X}}
\newcommand{\calY}{\mathcal{Y}}
\newcommand{\calZ}{\mathcal{Z}}

\newcommand{\sfA}{\mathsf{A}}
\newcommand{\sfB}{\mathsf{B}}
\newcommand{\sfC}{\mathsf{C}}

\newcommand{\sfF}{\mathsf{F}}

\newcommand{\sfM}{\mathsf{M}}
\newcommand{\sfN}{\mathsf{N}}

\newcommand{\sfP}{\mathsf{P}}

\newcommand{\sfR}{\mathsf{R}}
\newcommand{\sfS}{\mathsf{S}}
\newcommand{\sfT}{\mathsf{T}}

\newcommand{\sfV}{\mathsf{V}}
\newcommand{\sfW}{\mathsf{W}}
\newcommand{\sfX}{\mathsf{X}}
\newcommand{\sfY}{\mathsf{Y}}

\newcommand{\classfont}{\mathrm}

\newcommand{\NP}{\classfont{NP}}

\newcommand{\PSPACE}{\classfont{PSPACE}}

\newcommand{\IP}{\classfont{IP}}

\newcommand{\QIP}{\classfont{QIP}}

\newcommand{\AM}{\classfont{AM}}
\newcommand{\QMA}{\classfont{QMA}}

\newcommand{\QAM}{\classfont{QAM}}
\newcommand{\QMAM}{\classfont{QMAM}}

\newcommand{\QPZK}{\classfont{QPZK}}

\newcommand{\HVQPZK}{\classfont{HVQPZK}}
\newcommand{\SZK}{\classfont{SZK}}
\newcommand{\QSZK}{\classfont{QSZK}}
\newcommand{\NISZK}{\classfont{NISZK}}

\newcommand{\HVSZK}{\classfont{HVSZK}}
\newcommand{\HVQSZK}{\classfont{HVQSZK}}
\newcommand{\ZK}{\classfont{ZK}}
\newcommand{\QZK}{\classfont{QZK}}

\newcommand{\HVQZK}{\classfont{HVQZK}}

\newcommand{\tr}{\mathrm{tr}}

\newcommand{\init}{\mathrm{init}}

\newcommand{\acc}{\mathrm{acc}}
\newcommand{\rej}{\mathrm{rej}}

\newcommand{\view}{\mathrm{view}}
\newcommand{\FAIL}{\mathsf{FAIL}}

\newcommand{\fail}{\mathrm{fail}}

\newcommand{\bra}[1]{\langle #1 \vert}

\newcommand{\ket}[1]{\vert #1 \rangle}

\newcommand{\ketbra}[1]{\vert #1 \rangle \langle #1 \vert}

\newcommand{\conjugate}[1]{#1^{\dagger}}

\newcommand{\norm}[1]{\Vert #1 \Vert}

\newcommand{\trnorm}[1]{\Vert #1 \Vert_{\mathrm{tr}}}

\newcommand{\diamondnorm}[1]{\Vert #1 \Vert_{\diamond}}

\newcommand{\abs}[1]{\vert #1 \vert}
\newcommand{\absL}[1]{\left\vert #1 \right\vert}

\newcommand{\ceil}[1]{\lceil #1 \rceil}

\newcommand{\bigceil}[1]{\bigl\lceil #1 \bigr\rceil}

\newcommand{\floor}[1]{\lfloor #1 \rfloor}
\newcommand{\floorL}[1]{\left\lfloor #1 \right\rfloor}

\newcommand{\lrangle}[1]{\langle #1 \rangle}

\newcommand{\function}[3]{{#1 \colon #2 \rightarrow #3}}
\newcommand{\set}[2]{{\{ #1 \colon #2 \}}}

\newcommand{\bigset}[2]{{\bigl\{ #1 \colon #2 \bigr\}}}

\newcommand{\yes}{\mathrm{yes}}
\newcommand{\no}{\mathrm{no}}

\newcommand{\Natural}{\mathbb{N}}
\newcommand{\Integers}{\mathbb{Z}}
\newcommand{\Nonnegative}{\Integers^{+}}

\newcommand{\Binary}{{\{ 0, 1 \}}}
\newcommand{\Density}{\mathbf{D}}

\newcommand{\Admissible}{\mathbf{T}}
\newcommand{\Unitary}{\mathbf{U}}

\begin{document}

\sloppy


\title{\Large
  \textbf{
    General Properties of Quantum Zero-Knowledge Proofs
  }\\
}

\author{
  Hirotada Kobayashi\\
  \texttt{hirotada@nii.ac.jp}\\
  [2mm]
  Principles of Informatics Research Division\\
  National Institute of Informatics\\
  2-1-2 Hitotsubashi, Chiyoda-ku, Tokyo 101-8430, Japan
}

\date{8 May 2007}

\maketitle
\thispagestyle{empty}
\pagestyle{plain}
\setcounter{page}{0}


\begin{abstract}
This paper studies the complexity classes $\QZK$ and $\HVQZK$,
the classes of problems
having a quantum computational zero-knowledge proof system
and an \emph{honest-verifier} quantum computational zero-knowledge proof system,
respectively.
The results proved in this paper include:
\begin{itemize}
\item
  ${\HVQZK = \QZK}$.
\item
  Any problem in $\QZK$
  has a \emph{public-coin} quantum computational zero-knowledge proof system.
\item
  Any problem in $\QZK$
  has a quantum computational zero-knowledge proof system
  of \emph{perfect completeness}.
\item
  Any problem in $\QZK$
  has a \emph{three-message public-coin} quantum computational zero-knowledge proof system
  of perfect completeness
  with polynomially small error in soundness
  (hence with arbitrarily small constant error in soundness).
\end{itemize}
All the results proved in this paper are unconditional,
i.e., they do not rely any computational assumptions
such as the existence of quantum one-way functions or permutations.
For the classes $\QPZK$, $\HVQPZK$, and $\QSZK$ 
of problems
having a quantum perfect zero-knowledge proof system,
an honest-verifier quantum perfect zero-knowledge proof system,
and a quantum statistical zero-knowledge proof system,
respectively,
the following new properties are proved:
\begin{itemize}
\item
  ${\HVQPZK = \QPZK}$.
\item
  Any problem in $\QPZK$
  has a \emph{public-coin} quantum perfect zero-knowledge proof system.
\item
  Any problem in $\QSZK$
  has a quantum statistical zero-knowledge proof system
  of \emph{perfect completeness}.
\item
  Any problem in $\QSZK$
  has a \emph{three-message public-coin} quantum statistical zero-knowledge proof system
  of perfect completeness
  with polynomially small error in soundness
  (hence with arbitrarily small constant error in soundness).
\end{itemize}
It is stressed that the proofs for all the statements are direct
and do not use complete promise problems or those equivalents.
This gives a \emph{unified framework}
that works well for all of
quantum perfect, statistical, and computational zero-knowledge proofs.
In particular, this enables us to prove properties
even on the computational and perfect zero-knowledge proofs
for which no complete promise problems nor those equivalents are known.
\end{abstract}

\clearpage


\section{Introduction}
\label{Section: introduction}


\subsection{Background}
\label{Subsection: background}

Zero-knowledge proof systems
were introduced by Goldwasser,~Micali,~and~Rackoff~\cite{GolMicRac89SIComp},
and have played a central role in modern cryptography since then.
Intuitively, an interactive proof system is zero-knowledge
if \emph{any} verifier who communicates with the \emph{honest} prover
learns nothing except for the validity of the statement being proved
in that system.
By ``learns nothing'' we mean that
there exists a polynomial-time \emph{simulator}
whose output is indistinguishable from the output of the verifier
after communicating with the honest prover.
Depending on the strength of this indistinguishability,
several variants of zero-knowledge proofs have been investigated:
\emph{perfect} zero-knowledge
in which the output of the simulator is identical
to that of the verifier,
\emph{statistical} zero-knowledge
in which the output of the simulator is statistically close
to that of the verifier,
and
\emph{computational} zero-knowledge
in which the output of the simulator is
indistinguishable from that of the verifier
in polynomial time.
The most striking result on zero-knowledge proofs
would be that every problem in $\NP$
has a computational zero-knowledge proof system
under certain intractability assumptions~\cite{GolMicWig91JACM}
like the existence of one-way functions~\cite{Nao91JCrypto, HasImpLevLub99SIComp}.
It is also known that some problems have perfect or statistical zero-knowledge proof systems.
Among others,
the \textsc{Graph Isomorphism} problem
has a perfect zero-knowledge proof system~\cite{GolMicWig91JACM},
and some lattice problems
have statistical zero-knowledge proof systems~\cite{GolGol00JCSS}.

Another direction of studies on zero-knowledge proofs
has been to prove general properties of zero-knowledge proofs.
Sahai~and~Vadhan~\cite{SahVad03JACM} were the first
that took an approach of characterizing zero-knowledge proofs
by complete promise problems.
They showed that the \textsc{Statistical Difference} problem is complete
for the class $\HVSZK$ of problems having
an \emph{honest-verifier} statistical zero-knowledge proof system.
Here, the honest-verifier zero-knowledge is
a weaker notion of zero-knowledge
in which now zero-knowledge property holds
only against the \emph{honest} verifier who follows the specified protocol.
Using this complete promise problem,
they proved a number of general properties of $\HVSZK$
and simplified the proofs of several previously known results
including that $\HVSZK$ is in $\AM$~\cite{For89RC, AieHas91JCSS},
that $\HVSZK$ is closed under complement~\cite{Oka00JCSS},
and that any problem in $\HVSZK$
has a public-coin honest-verifier statistical zero-knowledge proof system~\cite{Oka00JCSS}.
Goldreich~and~Vadhan~\cite{GolVad99CCC} presented
another complete promise problem for $\HVSZK$,
called the \textsc{Entropy Difference} problem,
and obtained further properties of $\HVSZK$.
Since Goldreich,~Sahai,~and~Vadhan~\cite{GolSahVad98STOC} proved that
${\HVSZK = \SZK}$,
where $\SZK$ denotes
the class of problems having a statistical zero-knowledge proof system,
all the properties for $\HVSZK$ are inherited to $\SZK$
(except for those related to round complexity).
Along this line,
Goldreich,~Sahai,~and~Vadhan~\cite{GolSahVad99CRYPTO} gave
two complete promise problems
for the class $\NISZK$ of problems having
a non-interactive statistical zero-knowledge proof system,
and derived several properties of $\NISZK$.
More recently, Vadhan~\cite{Vad06SIComp} gave two characterizations,
the \textsc{Indistinguishability} characterization
and the \textsc{Conditional Pseudo-Entropy} characterization,
for the class $\ZK$ of problems having
a computational zero-knowledge proof system.
These are not complete promise problems,
but more or less analogous to complete promise problems
and play essentially same roles as complete promise problems
in his proof.
Using these characterizations,
Vadhan proved a number of general properties for $\ZK$ unconditionally
(i.e., not assuming any intractability assumptions),
such as that honest-verifier computational zero-knowledge
equals general computational zero-knowledge,
that public-coin computational zero-knowledge equals
general computational zero-knowledge,
and that computational zero-knowledge of perfect completeness
equals general two-sided bounded error computational zero-knowledge.

Quantum zero-knowledge proofs were first studied by Watrous~\cite{Wat02FOCS}
in a restricted situation of
\emph{honest-verifier} quantum statistical zero-knowledge proofs.
He gave an analogous characterization
to the classical case by Sahai~and~Vadhan~\cite{SahVad03JACM}
by showing that the \textsc{Quantum State Distinguishability} problem
is complete for the class $\HVQSZK$ of problems having
an honest-verifier quantum statistical zero-knowledge proof system.
Using this, he proved a number of general properties for $\HVQSZK$,
such as that $\HVQSZK$ is closed under complement,
that any problem in $\HVQSZK$
has a public-coin honest-verifier quantum statistical zero-knowledge proof system,
and that $\HVQSZK$ is in $\PSPACE$.
Very recently,
Ben-Aroya~and~Ta-Shma~\cite{BenTaS07quant-ph}
presented another complete promise problem for $\HVQSZK$,
called the \textsc{Quantum Entropy Difference} problem,
which is a quantum analogue of the result by
Goldreich~and~Vadhan~\cite{GolVad99CCC}.
Kobayashi~\cite{Kob03ISAAC} studied
non-interactive quantum perfect and statistical zero-knowledge proofs
again using a complete promise problem,
which can be viewed as a quantum version of the classical result
by Goldreich,~Sahai,~and~Vadhan~\cite{GolSahVad99CRYPTO}.
It has been a wide open problem
if there are nontrivial problems
that has a quantum zero-knowledge proof system
secure even against any dishonest quantum verifiers,
because of the difficulties arising from the ``rewinding'' technique~\cite{Gra97PhD},
which is commonly-used in classical zero-knowledge proofs.
Damg{\aa}rd, Fehr, and Salvail~\cite{DamFehSal04CRYPTO}
studied zero-knowledge proofs against dishonest quantum verifier,
but they assumed the restricted setting
of the common-reference-string model to avoid this rewinding problem.
Very recently, Watrous~\cite{Wat06STOC} settled this affirmatively.
He developed a quantum ``rewinding'' technique
by using a method that was originally developed in Ref.~\cite{MarWat05CC}
for the purpose of amplifying the success probability of $\QMA$,
a quantum version of $\NP$, without increasing
quantum witness sizes.
With this quantum rewinding technique, he proved that the classical protocol
for the \textsc{Graph Isomorphism} problem in Ref.~\cite{GolMicWig91JACM}
has a perfect zero-knowledge property
even against any dishonest \emph{quantum} verifiers,
and under some reasonable intractability assumption,
the classical protocol for $\NP$ in Ref.~\cite{GolMicWig91JACM}
has a computational zero-knowledge property
even against any dishonest \emph{quantum} verifiers.
He also proved that ${\HVQSZK = \QSZK}$,
where $\QSZK$ denotes
the class of problems having a quantum statistical zero-knowledge proof system.
This implies that
all the properties for $\HVQSZK$ proved in Ref.~\cite{Wat02FOCS}
are inherited to $\QSZK$ (except for those related to round complexity),
in particular, that
any problem in $\QSZK$
has a public-coin quantum statistical zero-knowledge proof system.


\subsection{Our Contribution}
\label{Subsection: contribution}

This paper proves a number of general properties
on quantum zero-knowledge proofs,
not restricted to quantum statistical zero-knowledge proofs.
Specifically, for quantum computational zero-knowledge proofs,
letting $\QZK$ and $\HVQZK$ denote
the classes of problems
having a quantum computational zero-knowledge proof system
and an \emph{honest-verifier} quantum computational zero-knowledge proof system,
respectively,
the following are proved among others:

\begin{theorem*}[Theorem~\ref{Theorem: HVQZK = QZK}]
${\HVQZK = \QZK}$.
\end{theorem*}

\begin{theorem*}[Theorem~\ref{Theorem: public-coin QZK = QZK}]
Any problem in $\QZK$
has a public-coin quantum computational zero-knowledge proof system.
\end{theorem*}

\begin{theorem*}[Theorem~\ref{Theorem: QZK = QZK with perfect completeness}]
Any problem in $\QZK$
has a quantum computational zero-knowledge proof system of perfect completeness.
\end{theorem*}

\begin{theorem*}[Theorem~\ref{Theorem: three-message public-coin QZK}]
Any problem in $\QZK$
has a three-message public-coin quantum computational zero-knowledge proof system
of perfect completeness
with soundness error probability at most $\frac{1}{p}$
for any polynomially bounded function
$\function{p}{\Nonnegative}{\Natural}$
(hence with arbitrarily small constant error in soundness).
\end{theorem*}

All the properties proved in this paper
on quantum computational zero-knowledge proofs
hold unconditionally,
meaning that they hold without any computational assumptions
such as the existence of quantum one-way functions or permutations.
Some of these properties may be regarded as quantum versions
of the results by Vadhan~\cite{Vad06SIComp}.
It is stressed, however, that our approach to prove these properties
is completely different from those the existing studies took
to prove general properties of classical or quantum zero-knowledge proofs.
No complete promise problems nor those equivalents are used in our proofs.
Instead, we \emph{directly} prove these properties,
which gives a \emph{unified framework}
that works well for all of
quantum perfect, statistical, and computational zero-knowledge proofs.

The idea is remarkably simple.
We start from any protocol of \emph{honest-verifier} quantum zero-knowledge,
and apply several modifications so that we finally obtain another protocol of honest-verifier quantum zero-knowledge
that possesses a number of desirable properties.
For instance, to prove that ${\HVQZK = \QZK}$,
we show that any protocol of honest-verifier quantum computational zero-knowledge
can be modified to another protocol of honest-verifier quantum computational zero-knowledge
(with some smaller gap between completeness and soundness accepting probabilities)
such that
(i) the protocol consists of three messages
and (ii) the protocol is public-coin
in which the message from the honest verifier consists of a single bit
that is an outcome of a classical fair coin-flipping.
Note that such modifications are possible
in the case of usual quantum interactive proofs~\cite{KitWat00STOC, MarWat05CC},
and we show that this is also the case
for \emph{honest-verifier} quantum computational zero-knowledge proofs.
Now we apply the quantum rewinding technique due to Watrous~\cite{Wat06STOC}
to show that the protocol is zero-knowledge
even against any \emph{dishonest} quantum verifiers.
The final tip is the sequential repetition,
which reduces completeness and soundness errors arbitrarily small.
This simultaneously shows
the equivalence of public-coin quantum computational zero-knowledge
and general quantum computational zero-knowledge.
To show that any quantum computational zero-knowledge proofs
can be made perfect complete,
now we have only to show that
any \emph{honest-verifier} quantum computational zero-knowledge proofs
can be made perfect complete.
Again a similar property is known to hold
for usual quantum interactive proofs~\cite{KitWat00STOC},
and we carefully modify the protocol
so that it holds
even for the honest-verifier quantum computational zero-knowledge case.
Using this modification as a preprocessing,
the previous argument shows
the equivalence of quantum computational zero-knowledge of perfect completeness
and general quantum computational zero-knowledge.
Combining all the desirable properties of
honest-verifier quantum computational zero-knowledge proofs
shown in this paper
with a careful application of the quantum rewinding technique,
we can show that
any problem in $\QZK$
has a three-message public-coin quantum computational zero-knowledge proof system
of perfect completeness
with soundness error at most $\frac{1}{p}$
for any polynomially bounded function $p$.

In fact, our approach above is very general and basically works well
even for quantum perfect and statistical zero-knowledge proofs.
In the quantum statistical zero-knowledge case,
all the properties shown for the quantum computational zero-knowledge case
also hold.
This gives alternative proofs of some of the properties
obtained in Refs.~\cite{Wat02FOCS, Wat06STOC},
and also shows the following new properties
of quantum statistical zero-knowledge proofs:

\begin{theorem*}[Theorem~\ref{Theorem: QSZK = QSZK with perfect completeness}]
Any problem in $\QSZK$
has a quantum statistical zero-knowledge proof system of perfect completeness.
\end{theorem*}

\begin{theorem*}[Theorem~\ref{Theorem: three-message public-coin QSZK}]
Any problem in $\QSZK$
has a three-message public-coin quantum statistical zero-knowledge proof system
of perfect completeness
with soundness error probability at most $\frac{1}{p}$
for any polynomially bounded function
$\function{p}{\Nonnegative}{\Natural}$
(hence with arbitrarily small constant error in soundness).
\end{theorem*}

In the quantum perfect zero-knowledge case,
however, not all the properties above can be shown to hold,
because very subtle points easily lose the \emph{perfect} zero-knowledge property.
In particular, our method of making protocols perfect complete
that works well for quantum computational and statistical zero-knowledge cases
no longer works well for quantum perfect zero-knowledge case.
Also, we need a very careful modification of the protocol
when parallelizing to three messages.
Still, we can show the following properties:

\begin{theorem*}[Theorem~\ref{Theorem: HVQPZK = QPZK}]
${\HVQPZK = \QPZK}$.
\end{theorem*}

\begin{theorem*}[Theorem~\ref{Theorem: public-coin QPZK = QPZK}]
Any problem in $\QPZK$
has a public-coin quantum perfect zero-knowledge proof system.
\end{theorem*}

\noindent
Note that
no such general properties are known
for the classical perfect zero-knowledge case.
As a bonus property, it is also proved that
the quantum perfect zero-knowledge
with a worst-case polynomial-time simulator
that is not allowed to output ``$\FAIL$''
is equivalent to
the one in which a simulator is allowed
to output ``$\FAIL$'' with small probability.
Again, such equivalence is not known in the classical case.


\subsection{Organization of This Paper}
\label{Subsection: organization}

This paper is organized as follows.
Section~\ref{Section: Preliminaries}
summarizes the notions and notations that are used in this paper.
Sections~\ref{Section: QPZK},~\ref{Section: QZK},~and~\ref{Section: QSZK}
treat our results for
quantum perfect, computational, and statistical zero-knowledge proofs, respectively.
In order to present a unified framework
that works well for all of
quantum perfect, computational, and statistical zero-knowledge proofs,
we first show the results for the perfect zero-knowledge case.
This may involve more careful modifications of the protocols
that are necessary only for the perfect zero-knowledge case,
but once we have presented how to modify the protocols,
we can avoid complications arising from imperfect zero-knowledge conditions
when proving zero-knowledge property,
which will be helpful to illustrate most of our proof structures in a simpler setting.
Section~\ref{Section: Equivalence of Two Definitions of Quantum Perfect Zero-Knowledge}
proves the equivalence of two different definitions of quantum perfect zero-knowledge.
Finally, Section~\ref{Section: conclusion}
concludes the paper with some open problems.


\section{Preliminaries}
\label{Section: Preliminaries}

We assume the reader is familiar with
classical zero-knowledge proof systems
and quantum interactive proof systems.
Detailed discussions of classical zero-knowledge proof systems
can be found in Refs.~\cite{Gol01Book, Gol02ECCC}, for instance,
while quantum interactive proof systems
are discussed in Refs.~\cite{Wat03TCS, KitWat00STOC, MarWat05CC}
and are reviewed in Appendix~\ref{Appendix: QIP model}.
We also assume familiarity with the quantum formalism,
including the quantum circuit model and
definitions of mixed quantum states,
admissible transformations (completely-positive trace-preserving mappings),
trace norm, diamond norm, and fidelity
(all of which are discussed in detail
in Refs.~\cite{NieChu00Book, KitSheVya02Book}, for instance).

Some of the notions and notations that are used in this paper
are summarized in this section.

Throughout this paper,
let $\Natural$ and $\Nonnegative$ denote
the sets of positive and nonnegative integers, respectively.
For every ${d \in \Natural}$,
let $I_d$ denote the identity operator of dimension $d$.
Also, for any Hilbert space $\calH$,
let $I_{\calH}$ denote the identity operator over $\calH$.
In this paper, all Hilbert spaces are of dimension power of two.


\subsection{Quantum Formalism}
\label{Subsection: Quantum Formalism}

For any Hilbert spaces $\calH$ and $\calK$,
let ${\Density(\calH)}$, ${\Unitary(\calH)}$, and ${\Admissible(\calH, \calK)}$
denote the sets of density operators over $\calH$,
unitary operators over $\calH$,
and admissible transformations from $\calH$ to $\calK$,
respectively.
For any Hilbert space $\calH$,
let $\ket{0_{\calH}}$ denote the quantum state in $\calH$
of which all the qubits are in state $\ket{0}$.

Let $\calH$ and $\calK$ be the Hilbert spaces
and let ${\Phi \in \Admissible(\calH, \calK)}$
be an admissible transformation.
Let $\calN$, $\calX$, and $\calY$ be Hilbert spaces
such that ${\calH \otimes \calX = \calK \otimes \calY = \calN}$.
A unitary transformation ${U_{\Phi} \in \Unitary(\calN)}$
is a \emph{unitary realization} of $\Phi$
if
${
  \tr_{\calY}
    U_{\Phi}
    \bigl(
      \rho \otimes \ketbra{0_{\calX}}
    \bigr)
    \conjugate{U_{\Phi}}
  =
  \Phi(\rho)
}$
for any ${\rho \in \Density(\calH)}$.

The following approximate version of unitary equivalence
is used in this paper.

\begin{lemma}[\cite{Wat02FOCS}]
For Hilbert spaces $\calH$ and $\calK$,
let ${\ket{\phi}, \ket{\psi} \in \calH \otimes \calK}$
satisfy that
${
  F(\tr_{\calK} \ketbra{\phi}, \tr_{\calK} \ketbra{\psi})
  \geq
  1 - \varepsilon
}$
for some ${\varepsilon \in [0,1]}$.
Then there exists a unitary transformation ${U \in \Unitary(\calK)}$
such that
${
  \norm{(I_{\calH} \otimes U) \ket{\phi} - \ket{\psi}}
  \leq
  \sqrt{2 \varepsilon}
}$.
\label{Lemma: approximate unitary equivalence}
\end{lemma}


\subsection{Quantum Circuits and Polynomial-Time Preparable Ensembles of Quantum States}
\label{Subsection: Quantum Circuits}

It is assumed that any quantum circuit $Q$ in this paper
is unitary and is composed of gates
in some reasonable, universal, finite set of unitary quantum gates.
For convenience,
we may identify a circuit $Q$ with the unitary operator it induces.

Since non-unitary and unitary quantum circuits
are equivalent in computational power~\cite{AhaKitNis98STOC},
it is sufficient to treat only unitary quantum circuits,
which justifies the above assumption.
For avoiding unnecessary complication, however,
the descriptions of procedures often include non-unitary operations
in the subsequent sections.
Even in such cases, it is always possible to construct
unitary quantum circuits
that essentially achieve the same procedures described.
A quantum circuit $Q$ is \emph{$q_\mathrm{in}$-in $q_\mathrm{out}$-out}
if it exactly implements a unitary realization $U_{\Phi}$
of some $q_\mathrm{in}$-in $q_\mathrm{out}$-out admissible transformation $\Phi$.
For convenience,
we may identify a circuit $Q$ with $\Phi$ in such a case.
As a special case of this,
a quantum circuit $Q$ is a \emph{generating circuit} of
a quantum state $\rho$ of $q$ qubits
if it exactly implements a unitary realization
of a zero-in $q$-out admissible transformation that always outputs $\rho$.

Following preceding studies
on quantum interactive and zero-knowledge proofs,
this paper uses the following notion of
polynomial-time uniformly generated families of quantum circuits.

A family ${\{ Q_x \}}$ of quantum circuits is
\emph{polynomial-time uniformly generated}
if there exists a deterministic procedure
that, on every input $x$, outputs a description of $Q_x$
and runs in time polynomial in $\abs{x}$.
It is assumed that the number of gates in any circuit
is not more than the length of the description of that circuit.
Hence $Q_x$ must have size polynomial in $\abs{x}$.

When proving statements concerning quantum perfect zero-knowledge proofs
or proofs having perfect completeness,
we assume that our universal gate set satisfies some conditions,
since these ``perfect'' properties may not hold
with an arbitrary universal gate set.
In fact,
this is also the case for some previous studies
on quantum interactive or zero-knowledge proofs,
including the papers by Kitaev~and~Watrous~\cite{KitWat00STOC}
and by Marriott~and~Watrous~\cite{MarWat05CC},
when deriving statements with perfect completeness property.
The correctness of our results
concerning quantum perfect zero-knowledge proofs
or proofs having perfect completeness
may be discussed under a similar assumption to those studies
on the choice of the universal gate set.
Fortunately, the author learned from John Watrous~\cite{Wat07private}
that the choice of the gate set would not be so critical
and all the ``perfect'' properties
claimed in Refs.~\cite{KitWat00STOC, MarWat05CC}
and in this paper
hold with any gate set such that
the Hadamard transformation and any classical reversible transformations
are exactly implementable.
Note that this condition is satisfied
by most of the standard gate sets
including the Shor basis~\cite{Sho96FOCS} consisting of
the Hadamard gate, the controlled-$i$-phase-shift gate, and the Toffoli gate.
These subtle issues regarding choices of the universal gate set
will be explained in detail
in Appendix~\ref{Appendix: Note on the Choice of Universal Gate Set}.
It is stressed, however, that all of our statements
not concerning quantum perfect zero-knowledge proofs
nor proofs having perfect completeness
do hold for an arbitrary choice of the universal gate set
(the completeness and soundness conditions
may become worse by negligible amounts in some of the claims,
which does not matter for the final main statements).

Finally,
this paper uses the following notion of
polynomial-time preparable ensembles of quantum states,
which was introduced in Ref.~\cite{Wat02FOCS}.

An ensemble ${\{ \rho_x \}}$ of quantum states is
\emph{polynomial-time preparable}
if there exists
a polynomial-time uniformly generated family ${\{ Q_x \}}$ of quantum circuits
such that each $Q_x$ is a generating circuit of $\rho_x$.
In what follows, we may use the notation ${\{ \rho(x) \}}$
instead of ${\{ \rho_x \}}$ for ensembles of quantum states
simply for descriptional convenience.


\subsection{Quantum Computational Indistinguishability}
\label{Subsection: Quantum Computational Indistinguishability}

We use the notions of quantum computational indistinguishability
introduced by Watrous~\cite{Wat06STOC}:
polynomially quantum indistinguishable ensembles of quantum states
and
polynomially quantum indistinguishable ensembles of admissible transformations.

First, the quantum computational indistinguishability
between two ensembles of quantum states
is defined as follows.

\begin{definition}
Let ${S \subseteq \Binary^{\ast}}$ be an infinite set
and let $\function{m}{\Nonnegative}{\Natural}$
be a polynomially bounded function.
For each ${x \in S}$,
let $\rho_x$ and $\sigma_x$ be mixed states of ${m(\abs{x})}$ qubits.
The ensembles $\set{\rho_x}{x \in S}$ and $\set{\sigma_x}{x \in S}$
are \emph{polynomially quantum indistinguishable}
if, for every choice of
\begin{itemize}
\item
  polynomially bounded functions $\function{k,p,s}{\Nonnegative}{\Natural}$,
\item
  an ensemble $\set{\xi_x}{x \in S}$,
  where $\xi_x$ is a mixed state of ${k(\abs{x})}$ qubits,
  and
\item
  an ${(m(\abs{x}) + k(\abs{x}))}$-in $1$-out quantum circuit $Q$
  of size at most ${s(\abs{x})}$,
\end{itemize}
it holds that
\[
\absL{
  \bra{1} Q (\rho_x \otimes \xi_x) \ket{1}
  -
  \bra{1} Q (\sigma_x \otimes \xi_x) \ket{1}
}
<
\frac{1}{p(\abs{x})}
\]
for all but finitely many ${x \in S}$.
\end{definition}

Next, the quantum computational indistinguishability
between two ensembles of admissible transformations
is defined as follows.

\begin{definition}
Let ${S \subseteq \Binary^{\ast}}$ be an infinite set
and let $\function{l,m}{\Nonnegative}{\Natural}$
be polynomially bounded functions.
For each ${x \in S}$,
let $\Phi_x$ and $\Psi_x$ be
${l(\abs{x})}$-in ${m(\abs{x})}$-out admissible transformations.
The ensembles $\set{\Phi_x}{x \in S}$ and $\set{\Psi_x}{x \in S}$
are \emph{polynomially quantum indistinguishable}
if, for every choice of
\begin{itemize}
\item
  polynomially bounded functions $\function{k,p,s}{\Nonnegative}{\Natural}$,
\item
  an ensemble $\set{\xi_x}{x \in S}$,
  where $\xi_x$ is a mixed state of ${l(\abs{x}) + k(\abs{x})}$ qubits,
  and
\item
  an ${(m(\abs{x}) + k(\abs{x}))}$-in $1$-out quantum circuit $Q$
  of size at most ${s(\abs{x})}$,
\end{itemize}
it holds that
\[
\absL{
  \bra{1} Q \bigl( (\Phi_x \otimes I_{2^{k(\abs{x})}}) (\xi_x) \bigr) \ket{1}
  -
  \bra{1} Q \bigl( (\Psi_x \otimes I_{2^{k(\abs{x})}}) (\xi_x) \bigr) \ket{1}
}
<
\frac{1}{p(\abs{x})}
\]
for all but finitely many ${x \in S}$.
\end{definition}

In what follows, we will often use the term
``computationally indistinguishable''
instead of ``polynomially quantum indistinguishable''
for simplicity.
Also, we will often informally say that
mixed states $\rho_x$ and $\sigma_x$
or admissible transformations $\Phi_x$ and $\Psi_x$
are computationally indistinguishable when ${x \in S}$
to mean that the ensembles
$\set{\rho_x}{x \in S}$ and $\set{\sigma_x}{x \in S}$
or
$\set{\Phi_x}{x \in S}$ and $\set{\Psi_x}{x \in S}$
are polynomially quantum indistinguishable.


\subsection{Quantum Zero-Knowledge Proofs}
\label{Subsection: QZK definitions}

For readability, in what follows,
the arguments $x$ and $n$ are dropped in the various functions,
if it is not confusing.
It is assumed that operators acting on subsystems of a given system
are extended to the entire system by tensoring with the identity,
since it will be clear from context
upon what part of a system a given operator acts.
Although all the statements in this paper
can be proved only in terms of languages
without using promise problems~\cite{EveSelYac84InfoCont},
in what follows we define models and prove statements
in terms of promise problems,
for generality
and for the compatibility with some other studies
on quantum zero-knowledge proofs~\cite{Wat02FOCS, Kob03ISAAC, Wat06STOC, BenTaS07quant-ph}.

First we define the notions of
various \emph{honest-verifier} quantum zero-knowledge proofs
following a manner in Ref.~\cite{Wat02FOCS}
for the statistical zero-knowledge case.
Given a quantum verifier $V$ and a quantum prover $P$,
let ${\view_{V,P}(x,j)}$ be the quantum state
that $V$ possesses immediately after the $j$th transformation of $P$
during an execution of the protocol between $V$ and $P$.
In other words,
${\view_{V,P}(x,j)}$ is the state obtained by
tracing out the private space of $P$
from the state of the entire system
immediately after the $j$th transformation of $P$.

Now we define the classes
${\HVQPZK(m,c,s)}$, ${\HVQSZK(m,c,s)}$, and ${\HVQZK(m,c,s)}$
of problems
having $m$-message honest-verifier
quantum perfect, statistical, and computational
zero-knowledge proof systems, respectively,
with completeness accepting probability at least $c$
and soundness accepting probability at most $s$.

\begin{definition}
Given a polynomially bounded function
$\function{m}{\Nonnegative}{\Natural}$
and functions $\function{c, s}{\Nonnegative}{[0,1]}$,
a problem ${A = \{ A_{\yes}, A_{\no} \}}$ is in
${\HVQPZK(m,c,s)}$
iff there exists an $m$-message honest quantum verifier $V$
and an $m$-message honest quantum prover $P$
such that
\begin{description}
\item[\textnormal{(Completeness and Soundness)}]
 ${(V,P)}$ forms an $m$-message quantum interactive proof system
with completeness accepting probability at least $c$
and
soundness accepting probability at most $s$,
\item[\textnormal{(Honest-Verifier Perfect Zero-Knowledge)}]
there exists a polynomial-time preparable ensembles ${\{S_V(x,j)\}}$ of quantum states
such that
${S_V(x,j) = \view_{V,P}(x,j)}$
for every ${x \in A_{\yes}}$ and for each ${1 \leq j \leq \bigceil{\frac{m(\abs{x})}{2}}}$.
\end{description}
\label{Definition: HVQPZK(m,c,s)}
\end{definition}

\begin{definition}
Given a polynomially bounded function
$\function{m}{\Nonnegative}{\Natural}$
and functions $\function{c, s}{\Nonnegative}{[0,1]}$,
a problem ${A = \{ A_{\yes}, A_{\no} \}}$ is in
${\HVQSZK(m,c,s)}$
iff there exists an $m$-message honest quantum verifier $V$
and an $m$-message honest quantum prover $P$
such that
\begin{description}
\item[\textnormal{(Completeness and Soundness)}]
 ${(V,P)}$ forms an $m$-message quantum interactive proof system
with completeness accepting probability at least $c$
and
soundness accepting probability at most $s$,
\item[\textnormal{(Honest-Verifier Statistical Zero-Knowledge)}]
there exists a polynomial-time preparable ensembles ${\{S_V(x,j)\}}$ of quantum states
such that
$\trnorm{S_V(x,j) - \view_{V,P}(x,j)}$
is negligible with respect to $\abs{x}$
for all but finitely many
${(x,j) \in A_{\yes} \times \bigl\{1, \ldots, \bigceil{\frac{m(\abs{x})}{2}} \bigr\}}$.
\end{description}
\label{Definition: HVQSZK(m,c,s)}
\end{definition}

\begin{definition}
Given a polynomially bounded function
$\function{m}{\Nonnegative}{\Natural}$
and functions $\function{c, s}{\Nonnegative}{[0,1]}$,
a problem ${A = \{ A_{\yes}, A_{\no} \}}$ is in
${\HVQZK(m,c,s)}$
iff there exists an $m$-message honest quantum verifier $V$
and an $m$-message honest quantum prover $P$
such that
\begin{description}
\item[\textnormal{(Completeness and Soundness)}]
 ${(V,P)}$ forms an $m$-message quantum interactive proof system
with completeness accepting probability at least $c$
and
soundness accepting probability at most $s$,
\item[\textnormal{(Honest-Verifier Computational Zero-Knowledge)}]
there exists a polynomial-time preparable ensembles ${\{S_V(x,j)\}}$ of quantum states
such that
the ensembles $\bigset{S_V(x,j)}{x \in A_{\yes} \text{ and } j \in \bigl\{1, \ldots, \bigceil{\frac{m(\abs{x})}{2}} \bigr\}}$
and $\bigset{\view_{V,P}(x,j)}{x \in A_{\yes} \text{ and } j \in \bigl\{1, \ldots, \bigceil{\frac{m(\abs{x})}{2}} \bigr\}}$
are polynomially quantum indistinguishable.
\end{description}
\label{Definition: HVQZK(m,c,s)}
\end{definition}

\begin{remark}
In the original definition of
honest-verifier quantum statistical zero-knowledge by Watrous~\cite{Wat02FOCS},
the simulator is required to simulate
the quantum state that $V$ possesses immediately after the $j$th \emph{message},
for every $j$.
That is, regardless of whether the $j$th message is sent from $P$ or from $V$,
the simulator must be able to simulate
the quantum state that $V$ possesses immediately after the $j$th message.
In our definition,
the simulator is required to simulate it
only when the $j$th message is from $P$.
Notice, however, that
every transformation of $V$ is necessarily simulatable by the simulator,
which implies that our condition is sufficient
and does not weaken the honest-verifier zero-knowledge property.
\end{remark}

Using these,
we define the classes $\HVQPZK$, $\HVQSZK$, and $\HVQZK$ of problems
having honest-verifier quantum perfect, statistical, and computational
zero-knowledge proof systems, respectively.

\begin{definition}
A problem ${A = \{ A_{\yes}, A_{\no} \}}$ is in
$\HVQPZK$
if there exists a polynomially bounded function
$\function{m}{\Nonnegative}{\Natural}$
such that $A$ is in
${\HVQPZK \left(m, \frac{2}{3}, \frac{1}{3} \right)}$.
\label{Definition: HVQPZK}
\end{definition}

\begin{definition}
A problem ${A = \{ A_{\yes}, A_{\no} \}}$ is in
$\HVQSZK$
if there exists a polynomially bounded function
$\function{m}{\Nonnegative}{\Natural}$
such that $A$ is in
${\HVQSZK \left(m, \frac{2}{3}, \frac{1}{3} \right)}$.
\label{Definition: HVQSZK}
\end{definition}

\begin{definition}
A problem ${A = \{ A_{\yes}, A_{\no} \}}$ is in
$\HVQZK$
if there exists a polynomially bounded function
$\function{m}{\Nonnegative}{\Natural}$
such that $A$ is in
${\HVQZK \left(m, \frac{2}{3}, \frac{1}{3} \right)}$.
\label{Definition: HVQZK}
\end{definition}

Note that it is easy to see that we can amplify the success probability
of honest-verifier quantum perfect/statistical/computational
zero-knowledge proof systems
by a sequential repetition,
which justifies Definitions~\ref{Definition: HVQPZK},~\ref{Definition: HVQSZK},~and~\ref{Definition: HVQZK}.

Next we define the notions of various quantum zero-knowledge proofs
following a manner in Ref.~\cite{Wat06STOC}.

Let $V$ be an arbitrary quantum verifier.
Suppose that $V$ possesses some auxiliary quantum state in ${\Density(\calA)}$
at the beginning for some Hilbert space $\calA$,
and possesses some quantum state in ${\Density(\calZ)}$
after the protocol for some Hilbert space $\calZ$.
For such $V$, for any quantum prover $P$,
and for every ${x \in \Binary^{\ast}}$,
let ${\lrangle{V, P}(x)}$ denote the admissible transformation
in ${\Admissible(\calA, \calZ)}$
induced by the interaction between $V$ and $P$ on input $x$.
We call this ${\lrangle{V, P}(x)}$
the \emph{induced admissible transformation}
from $V$, $P$, and $x$.

We define the classes
${\QPZK(m,c,s)}$, ${\QSZK(m,c,s)}$, and ${\QZK(m,c,s)}$
of problems
having $m$-message quantum perfect, statistical, and computational
zero-knowledge proof systems, respectively,
with completeness accepting probability at least $c$
and soundness accepting probability at most $s$,
as follows.

\begin{definition}
Given a polynomially bounded function
$\function{m}{\Nonnegative}{\Natural}$
and functions $\function{c, s}{\Nonnegative}{[0,1]}$,
a problem ${A = \{ A_{\yes}, A_{\no} \}}$ is in
${\QPZK(m,c,s)}$
iff there exists an $m$-message honest quantum verifier $V$
and an $m$-message honest quantum prover $P$
such that
\begin{description}
\item[\textnormal{(Completeness and Soundness)}]
 ${(V,P)}$ forms an $m$-message quantum interactive proof system
with completeness accepting probability at least $c$
and
soundness accepting probability at most $s$,
\item[\textnormal{(Perfect Zero-Knowledge)}]
for any $m$-message quantum verifier $V'$,
there exists a polynomial-time uniformly generated family ${\{Q_x\}}$ of quantum circuits,
where each $Q_x$ exactly implements an admissible transformation $S_{V'}(x)$,
such that
${S_{V'}(x) = \lrangle{V',P}(x)}$
for every ${x \in A_{\yes}}$,
where ${\lrangle{V',P}(x)}$ is the induced admissible transformation
from $V'$, $P$, and $x$.
\end{description}
\label{Definition: QPZK(m,c,s)}
\end{definition}

\begin{definition}
Given a polynomially bounded function
$\function{m}{\Nonnegative}{\Natural}$
and functions $\function{c, s}{\Nonnegative}{[0,1]}$,
a problem ${A = \{ A_{\yes}, A_{\no} \}}$ is in
${\QSZK(m,c,s)}$
iff there exists an $m$-message honest quantum verifier $V$
and an $m$-message honest quantum prover $P$
such that
\begin{description}
\item[\textnormal{(Completeness and Soundness)}]
 ${(V,P)}$ forms an $m$-message quantum interactive proof system
with completeness accepting probability at least $c$
and
soundness accepting probability at most $s$,
\item[\textnormal{(Statistical Zero-Knowledge)}]
for any $m$-message quantum verifier $V'$,
there exists a polynomial-time uniformly generated family ${\{Q_x\}}$ of quantum circuits,
where each $Q_x$ exactly implements an admissible transformation $S_{V'}(x)$,
such that
$\diamondnorm{S_{V'}(x) - \lrangle{V',P}(x)}$
is negligible with respect to $\abs{x}$
for all but finitely many ${x \in A_{\yes}}$,
where ${\lrangle{V',P}(x)}$ is the induced admissible transformation
from $V'$, $P$, and $x$.
\end{description}
\label{Definition: QSZK(m,c,s)}
\end{definition}

\begin{definition}
Given a polynomially bounded function
$\function{m}{\Nonnegative}{\Natural}$
and functions $\function{c, s}{\Nonnegative}{[0,1]}$,
a problem ${A = \{ A_{\yes}, A_{\no} \}}$ is in
${\QZK(m,c,s)}$
iff there exists an $m$-message honest quantum verifier $V$
and an $m$-message honest quantum prover $P$
such that
\begin{description}
\item[\textnormal{(Completeness and Soundness)}]
 ${(V,P)}$ forms an $m$-message quantum interactive proof system
with completeness accepting probability at least $c$
and
soundness accepting probability at most $s$,
\item[\textnormal{(Computational Zero-Knowledge)}]
for any $m$-message quantum verifier $V'$,
there exists a polynomial-time uniformly generated family ${\{Q_x\}}$ of quantum circuits,
where each $Q_x$ exactly implements an admissible transformation $S_{V'}(x)$,
such that
the ensembles $\set{S_{V'}(x)}{x \in A_{\yes}}$
and $\set{\lrangle{V',P}(x)}{x \in A_{\yes}}$
are polynomially quantum indistinguishable,
where ${\lrangle{V',P}(x)}$ is the induced admissible transformation
from $V'$, $P$, and $x$.
\end{description}
\label{Definition: QZK(m,c,s)}
\end{definition}

Using these,
we define the classes $\QPZK$, $\QSZK$, and $\QZK$ of problems
having quantum perfect, statistical, and computational
zero-knowledge proof systems, respectively.

\begin{definition}
A problem ${A = \{ A_{\yes}, A_{\no} \}}$ is in
$\QPZK$
if there exists a polynomially bounded function
$\function{m}{\Nonnegative}{\Natural}$
such that $A$ is in
${\QPZK \left(m, \frac{2}{3}, \frac{1}{3} \right)}$.
\label{Definition: QPZK}
\end{definition}

\begin{definition}
A problem ${A = \{ A_{\yes}, A_{\no} \}}$ is in
$\QSZK$
if there exists a polynomially bounded function
$\function{m}{\Nonnegative}{\Natural}$
such that $A$ is in
${\QSZK \left(m, \frac{2}{3}, \frac{1}{3} \right)}$.
\label{Definition: QSZK}
\end{definition}

\begin{definition}
A problem ${A = \{ A_{\yes}, A_{\no} \}}$ is in
$\QZK$
if there exists a polynomially bounded function
$\function{m}{\Nonnegative}{\Natural}$
such that $A$ is in
${\QZK \left(m, \frac{2}{3}, \frac{1}{3} \right)}$.
\label{Definition: QZK}
\end{definition}

Note that again it is not hard to see that we can amplify the success probability
of quantum perfect/statistical/computational
zero-knowledge proof systems
by a sequential repetition,
which justifies Definitions~\ref{Definition: QPZK},~~\ref{Definition: QSZK},~and~\ref{Definition: QZK}.

\begin{remark}
It is noted that,
in the classical case,
the most common definition of perfect zero-knowledge proofs
seems to allow the simulator to output ``$\FAIL$''
with small probability,
say, with probability at most $\frac{1}{2}$~\cite{Gol01Book, SahVad03JACM}.
Adopting this convention
leads to alternative definitions of
honest-verifier and general quantum perfect zero-knowledge proof systems.
At a glance,
these two types of definitions
seem likely to form different complexity classes
of quantum perfect zero-knowledge proofs.
Fortunately, it is proved
from our results shown in Section~\ref{Section: QPZK} that
it is \emph{not} the case
and the two types of definitions
result in the same complexity class
of quantum perfect zero-knowledge proofs.
It is stressed that such equivalence is not known in the classical case.
See Section~\ref{Section: Equivalence of Two Definitions of Quantum Perfect Zero-Knowledge}
for further discussions on the definitions of
quantum perfect zero-knowledge.
\end{remark}


\section{Perfect Zero-Knowledge Case}
\label{Section: QPZK}

We first discuss the case of quantum perfect zero-knowledge proofs.
This gives a unified framework
that works well for all of
quantum perfect, statistical, and computational zero-knowledge proofs.
Although we need very careful modifications of the protocols
that are necessary only for the perfect zero-knowledge case,
once we have presented how to modify the protocols,
we can avoid complications arising from imperfect zero-knowledge conditions
when proving zero-knowledge property.
Indeed,
the cases of quantum computational and statistical zero-knowledge proofs
are proved in almost same ways, as will be discussed later,
except that we need bit more complicated arguments
when proving zero-knowledge conditions.


\subsection{Parallelization of Honest-Verifier Quantum Perfect Zero-Knowledge Proof Systems}
\label{Subsection: parallelization of HVQPZK}

This subsection proves that
any honest-verifier quantum perfect zero-knowledge proof system
that involves polynomially many messages
can be parallelized to one that involves only three messages.

In the case of usual quantum interactive proofs,
Kitaev~and~Watrous~\cite{KitWat00STOC} proved
the parallelizability to three messages.
Here we modify their method so that
it works well with honest-verifier quantum perfect zero-knowledge proofs.
Actually, the method due to Kitaev and Watrous
works well even in the cases of
honest-verifier quantum statistical or computational zero-knowledge proofs
(if the completeness error is negligible,
which may be assumed without loss of generality
since the success probability can be amplified by sequential repetition),
and thus, we do not need our modified version in these cases.
However, we do need our modified version
in the case of honest-verifier quantum perfect zero-knowledge proofs,
since the Kitaev-Watrous method
may not preserve the \emph{perfect} zero-knowledge property
for proof systems of imperfect completeness.
We explain this in more detail.

The main idea in the original parallelization protocol
in Ref.~\cite{KitWat00STOC}
is that the verifier receives each snapshot state of the underlying protocol
as the first message,
and then checks if the following three properties are satisfied:
(i) the first snapshot state is a legal state
in the underlying protocol after the first message,
(ii) the last snapshot state can make the original verifier accept,
and (iii) any two consecutive snapshot states
are indeed transformable with each other by one round of communication.
In order to check these three,
at the first transformation of the verifier
in the original parallelization protocol in Ref.~\cite{KitWat00STOC},
he first checks if the conditions (i) and (ii) really hold
for the received snapshot states,
which aims to prevent a dishonest prover from preparing
any illegal sequence of snapshot states
that can pass the check for the condition (iii)
by violating the conditions on the initial and last snapshot states.
The problem arises here, in the check for the last snapshot state,
when we want to parallelize a protocol of honest-verifier quantum perfect zero-knowledge
with imperfect completeness.
Because of imperfect completeness,
the verifier's check can fail even if the honest prover
prepares every snapshot state honestly,
which means that the verifier's check causes a small perturbation
to the snapshot states.
Now we have difficulty in \emph{perfectly} simulating the behavior of the honest prover
with respect to this perturbed state,
which causes the loss of the perfect zero-knowledge property.

To avoid this difficulty,
we modify the parallelization protocol as follows.
Our basic idea is to postpone the verifier's check for the last snapshot state
until after the third message.
At the final verification of the verifier,
he either carries out the postponed check for the last snapshot state
with probability $\frac{1}{2}$,
or just carries out the original final verification procedure
with probability $\frac{1}{2}$.
Now the honest-verifier perfect zero-knowledge property
becomes straightforward,
since there is no perturbation to all the snapshot states
until after the last transformation of the verifier.
The completeness property cannot become worse
than that in the original protocol.
However, the soundness condition now becomes a bit harder to prove,
because we can no longer assume that
a sequence of snapshot states prepared by a dishonest prover
satisfies the condition (ii),
when analyzing the probability to pass the transformability test for (iii).
To overcome this,
we show a general property in quantum information theory
in Lemma~\ref{Lemma: sum of sqrt of fidelity},
which is a generalization of Lemma~5~in~Ref.~\cite{KitWat00STOC}.
This generalization enables us
to analyze the case in which
the last snapshot state may not necessarily
make the original verifier accept,
and thus, has much more flexibility than Lemma~5~in~Ref.~\cite{KitWat00STOC},
which is applicable only to the case
in which the last snapshot state makes the original verifier accept
with certainty.

\begin{lemma}
Let $\calV$ and $\calM$ be any Hilbert spaces.
For a positive integer ${k \geq 2}$ and ${\varepsilon, \delta \in [0,1]}$
such that ${\varepsilon < \delta}$,
suppose that a sequence of unitary operators
${V_1, \ldots, V_k \in \Unitary(\calV \otimes \calM)}$
and a projection operator $\Pi$ acting over ${\calV \otimes \calM}$
onto some subspace of ${\calV \otimes \calM}$
satisfy that
${
  \norm{
    \Pi V_k P_{k-1} V_{k-1} \cdots P_1 V_1
    \ket{0_{\calV \otimes \calM \otimes \calP}}
  }^2
  \leq
  1 - \delta
}$
for any Hilbert space $\calP$
and any sequence of unitary operators
${P_1, \ldots, P_{k-1} \in \Unitary(\calM \otimes \calP)}$.
Then, for any sequence
${\rho_1, \ldots, \rho_k \in \Density(\calV \otimes \calM)}$
such that ${\rho_1 = \ketbra{0_{\calV \otimes \calM}}}$
and ${\tr \Pi V_k \rho_k \conjugate{V_k} \geq 1 - \varepsilon}$,
\[
\sum_{j=1}^{k-1}
  F(\tr_{\calM} V_j \rho_j \conjugate{V_j}, \tr_{\calM} \rho_{j+1})
\leq
(k-1) - \frac{(\sqrt{1 - \varepsilon} - \sqrt{1 - \delta})^2}{2(k-1)}.
\]
\label{Lemma: sum of sqrt of fidelity}
\end{lemma}

\begin{proof}
Let $\calP$ be a sufficiently large Hilbert space
so that we can take a purification
${\ket{\psi_j} \in \calV \otimes \calM \otimes \calP}$
of $\rho_j$ for each ${2 \leq j \leq k-1}$,
and let ${\ket{\psi_1} = \ket{0_{\calV \otimes \calM \otimes \calP}}}$.
Notice that $\ket{\psi_1}$ is a purification of $\rho_1$,
and ${V_j \ket{\psi_j}}$ is a purification of ${V_j \rho_j \conjugate{V_j}}$,
for each ${1 \leq j \leq k-1}$.
Let
${\Delta_j = 1 - F(\tr_{\calM} V_j \rho_j \conjugate{V_j}, \tr_{\calM} \rho_{j+1})}$
for each ${1 \leq j \leq k-1}$.
It follows from Lemma~\ref{Lemma: approximate unitary equivalence}
that there exists a unitary transformation
${P_j \in \Unitary(\calM \otimes \calP)}$
such that
${
  \norm{\ket{\psi_{j+1}} - P_j V_j \ket{\psi_j}}
  \leq
  \sqrt{2 \Delta_j}
}$,
for each ${1 \leq j \leq k-1}$.
Hence we have
\[
\begin{split}
\hspace{1cm}&\hspace{-1cm}
\norm{
  \Pi V_k \ket{\psi_k}
  -
  \Pi V_k P_{k-1} V_{k-1} \cdots P_1 V_1 \ket{\psi_1}
}
\\
&
\leq
\norm{V_k \ket{\psi_k} - V_k P_{k-1} V_{k-1} \cdots P_1 V_1 \ket{\psi_1}}
\\
&
=
\norm{\ket{\psi_k} - P_{k-1} V_{k-1} \cdots P_1 V_1 \ket{\psi_1}}
\\
&
\leq
\norm{
  \ket{\psi_k}
  -
  P_{k-1} V_{k-1} \ket{\psi_{k-1}}
}
+
\sum_{j=1}^{k-2}
\norm{
  P_{k-1} V_{k-1} \cdots P_{j+1} V_{j+1} \ket{\psi_{j+1}}
  -
  P_{k-1} V_{k-1} \cdots P_j V_j \ket{\psi_j}
}
\\
&
=
\sum_{j=1}^{k-1}
\norm{
  \ket{\psi_{j+1}}
  -
  P_j V_j \ket{\psi_j}
}
\\
&
\leq
\sum_{j=1}^{k-1}
\sqrt{2 \Delta_j}.
\end{split}
\]
On the other hand,
\[
\begin{split}
\norm{\Pi V_k \ket{\psi_k}}
&
\leq
\norm{
  \Pi V_k \ket{\psi_k}
  -
  \Pi V_k P_{k-1} V_{k-1} \cdots P_1 V_1 \ket{\psi_1}
}
+
\norm{
  \Pi V_k P_{k-1} V_{k-1} \cdots P_1 V_1 \ket{\psi_1}
}
\\
&
\leq
\sum_{j=1}^{k-1}
\sqrt{2 \Delta_j}
+
\sqrt{1 - \delta}.
\end{split}
\]
Notice that
${\norm{\Pi V_k \ket{\psi_k}} \geq \sqrt{1 - \varepsilon}}$,
since $\ket{\psi_k}$ is a purification of $\rho_k$
and ${\tr \Pi V_k \rho_k \conjugate{V_k} \geq 1 - \varepsilon}$.
Therefore,
\[
\sum_{j=1}^{k-1} \sqrt{\Delta_j}
\geq
\frac{\sqrt{1 - \varepsilon} - \sqrt{1 - \delta}}{\sqrt{2}},
\]
and thus,
\[
\sum_{j=1}^{k-1}
  F(\tr_{\calM} V_j \rho_j \conjugate{V_j}, \tr_{\calM} \rho_{j+1})
=
\sum_{j=1}^{k-1} (1 - \Delta_j)
=
(k-1) - \sum_{j=1}^{k-1} \Delta_j
\leq
(k-1) - \frac{(\sqrt{1 - \varepsilon} - \sqrt{1 - \delta})^2}{2(k-1)},
\]
as desired.
\end{proof}

Using Lemma~\ref{Lemma: sum of sqrt of fidelity},
we can show that our modified parallelization protocol above indeed works well,
and we have the following lemma.

\begin{lemma}
Let $\function{m}{\Nonnegative}{\Natural}$
be a polynomially bounded function
and let
$\function{\varepsilon, \delta}{\Nonnegative}{[0,1]}$
be any functions
such that ${m \geq 4}$ and ${\varepsilon < \frac{\delta^2}{16(m+1)^2}}$.
Then,
${
  \HVQPZK(m, 1 - \varepsilon, 1 - \delta)
  \subseteq
  \HVQPZK \left( 3, 1 - \frac{\varepsilon}{2}, 1 - \frac{\delta^2}{32(m+1)^2} \right)
}$.
\label{Lemma: reducing number of messages of HVQPZK to three -- modified KW}
\end{lemma}

\begin{proof}
Let ${A = \{A_{\yes}, A_{\no}\}}$ be a problem
in ${\HVQPZK(m, 1 - \varepsilon, 1 - \delta)}$
and let $V$ be the corresponding $m$-message honest quantum verifier.
For simplicity,
it is assumed that $m$ takes only even values
(if ${m(n)}$ is odd for some ${n \in \Nonnegative}$,
 we modify the protocol so that the verifier sends
 a ``dummy'' message to a prover as the first message
 when the input has length $n$ such that ${m(n)}$ is odd).
Let $\sfV$ be the quantum register
consisting of all the qubits in the private space of $V$,
and let $\sfM$ be that
consisting of all the qubits in the message channel between $V$ and the prover.
For every input $x$,
$V$ applies $V_j$ for his $j$th transformation
to the qubits in ${(\sfV, \sfM)}$
for ${1 \leq j \leq \frac{m}{2} + 1}$,
and performs the measurement ${\Pi = \{ \Pi_{\acc}, \Pi_{\rej} \}}$
at the end of the original protocol to decide acceptance or rejection.
We construct a protocol of a three-message honest quantum verifier $W$.

For every input $x$,
at the first message the new verifier $W$ receives quantum registers
$\sfV_j$ and $\sfM_j$ from the prover,
for ${2 \leq j \leq \frac{m}{2} + 1}$,
where each $\sfV_j$ and $\sfM_j$ consist of
the same number of qubits as $\sfV$ and $\sfM$, respectively.
$W$ expects that the qubits in ${(\sfV_j, \sfM_j)}$
form the quantum state
the original $m$-message verifier $V$ would possess
just after the ${2(j-1)}$-st message
(i.e., just before the $j$th transformation of the verifier)
of the original protocol,
for ${2 \leq j \leq \frac{m}{2} + 1}$.

Now $W$ prepares quantum registers $\sfV_1$ and $\sfM_1$,
which consist of the same number of qubits as $\sfV$ and $\sfM$, respectively,
and also prepares single-qubit quantum registers $\sfX$ and $\sfY$.
$W$ initializes all the qubits in $\sfV_1$ and $\sfM_1$ to state $\ket{0}$,
while prepares
${\ket{\Phi^+} = \frac{1}{\sqrt{2}}(\ket{0}\ket{0} + \ket{1}\ket{1})}$
in ${(\sfX, \sfY)}$.
$W$ then chooses ${r \in \bigl\{1, \ldots, \frac{m}{2} \bigr\}}$
uniformly at random,
applies $V_r$ to the qubits in ${(\sfV_r, \sfM_r)}$,
and sends $\sfY$ and $\sfM_r$ together with $r$ to the prover.

At the third message,
$W$ receives the quantum registers $\sfY$ and $\sfM_r$ from the prover.
Now $W$ chooses ${b \in \Binary}$ uniformly at random.
If ${b=0}$,
$W$ applies $V_{\frac{m}{2}+1}$ to the qubits in
${(\sfV_{\frac{m}{2}+1}, \sfM_{\frac{m}{2}+1})}$,
and accepts if and only if
the content of ${(\sfV_{\frac{m}{2}+1}, \sfM_{\frac{m}{2}+1})}$
corresponds to an accepting state in the original protocol.
On the other hand,
if ${b=1}$,
$W$ first performs a controlled-swap between
${(\sfV_r, \sfM_r)}$ and ${(\sfV_{r+1}, \sfM_{r+1})}$
using the qubit in $\sfX$ as the control,
then performs a controlled-not over the qubits in ${(\sfX, \sfY)}$
again using the qubit in $\sfX$ as the control,
and finally applies the Hadamard transformation to the qubit in $\sfX$.
$W$ accepts if and only if the qubit in $\sfX$ is in state $\ket{0}$.

The precise description of the protocol of $W$ is found
in Figure~\ref{Figure: parallelizing HVQPZK to three messages}.

\begin{figure}
\begin{algorithm*}{Honest Verifier's Three-Message Protocol}
\begin{step}
\item
  Receive quantum registers $\sfV_j$ and $\sfM_j$ from the prover,
  for ${2 \leq j \leq \frac{m}{2} + 1}$.%
  \label{first message}
\item
  Prepare quantum registers $\sfV_1$ and $\sfM_1$
  and single-qubit quantum registers $\sfX$ and $\sfY$.
  Initialize all the qubits in $\sfV_1$ and $\sfM_1$ to state $\ket{0}$,
  and prepare
  ${\ket{\Phi^+} = \frac{1}{\sqrt{2}}(\ket{0}\ket{0} + \ket{1}\ket{1})}$
  in ${(\sfX, \sfY)}$.
  Choose ${r \in \bigl\{1, \ldots, \frac{m}{2} \bigr\}}$
  uniformly at random
  and apply $V_r$ to the qubits in ${(\sfV_r, \sfM_r)}$.
  Send $\sfY$ and $\sfM_r$ together with $r$ to the prover.%
  \label{verifier's question}
\item
  Receive the quantum registers $\sfY$ and $\sfM_r$ from the prover.
  Choose ${b \in \Binary}$ uniformly at random.
  \begin{step}
  \item
    If ${b=0}$, do the following:\\
    Apply $V_{\frac{m}{2}+1}$ to the qubits in
    ${(\sfV_{\frac{m}{2}+1}, \sfM_{\frac{m}{2}+1})}$.
    Accept if
    the content of ${(\sfV_{\frac{m}{2}+1}, \sfM_{\frac{m}{2}+1})}$
    corresponds to an accepting state in the original protocol,
    and reject otherwise.
  \item
    If ${b=1}$, do the following:\\
    Perform a controlled-swap between
    ${(\sfV_r, \sfM_r)}$ and ${(\sfV_{r+1}, \sfM_{r+1})}$
    using the qubit in $\sfX$ as the control,
    and then perform a controlled-not over the qubits in ${(\sfX, \sfY)}$
    again using the qubit in $\sfX$ as the control.
    Apply the Hadamard transformation to the qubit in $\sfX$.
    Accept if the qubit in $\sfX$ is in state $\ket{0}$,
    and reject otherwise.
  \end{step}%
  \label{final verification step}
\end{step}
\end{algorithm*}
\caption{Honest verifier's three-message protocol.}
\label{Figure: parallelizing HVQPZK to three messages}
\end{figure}

For the completeness, suppose that the input $x$ is in $A_{\yes}$.

Let $P$ be the $m$-message honest quantum prover
for the original proof system,
and let $\sfP$ be the quantum register
consisting of all the qubits in the private space of $P$.
Denote by $\calV$, $\calM$, and $\calP$
the Hilbert spaces corresponding to the registers $\sfV$, $\sfM$, and $\sfP$,
respectively.
Let ${\ket{\psi_1} = \ket{0_{\calV \otimes \calM \otimes \calP}}}$
be the quantum state in ${(\sfV, \sfM, \sfP)}$,
and let ${\ket{\psi_j} \in \calV \otimes \calM \otimes \calP}$
be the quantum state in ${(\sfV, \sfM, \sfP)}$
just after the ${2(j-1)}$-st message
(i.e., just before the $j$th transformation of the verifier)
of the original protocol
if $V$ communicates with $P$ on input $x$,
for ${2 \leq j \leq \frac{m}{2} + 1}$.

Let $R$ be the honest quantum prover
in the constructed three-message system.
In addition to the registers $\sfV_j$ and $\sfM_j$,
$R$ prepares the quantum register $\sfP_j$ in his private space,
for ${1 \leq j \leq \frac{m}{2} + 1}$,
where each $\sfP_j$ consists of the same number of qubits as $\sfP$.
$R$ prepares $\ket{0_{\calP}}$ in $\sfP_1$
so that the qubits in ${(\sfV_1, \sfM_1, \sfP_1)}$ form $\ket{\psi_1}$.
At the first message of the constructed protocol,
$R$ generates $\ket{\psi_j}$ in ${(\sfV_j, \sfM_j, \sfP_j)}$,
and sends $\sfV_j$ and $\sfM_j$ to $W$,
for each ${2 \leq j \leq \frac{m}{2} + 1}$.

At the third message,
if $R$ receives $r$ together with the registers $\sfY$ and $\sfM_r$,
$R$ applies $P_r$ to the qubits in ${(\sfM_r, \sfP_r)}$,
where $P_j$ is the $j$th transformation of the original prover $P$
for each ${1 \leq j \leq \frac{m}{2}}$,
and then performs a controlled-swap between $\sfP_r$ and $\sfP_{r+1}$
using the qubit in $\sfY$ as the control.
$R$ then sends $\sfY$ and $\sfM_r$ back to $W$.

It is obvious that
$R$ can convince $W$ with probability at least ${1 - \varepsilon}$
if ${b=0}$ is chosen by $W$ at Step~\ref{final verification step},
since the qubits in ${(\sfV_{\frac{m}{2}+1}, \sfM_{\frac{m}{2}+1})}$
form the quantum state ${\tr_{\calP} \ketbra{\psi_{\frac{m}{2}+1}}}$.
From the construction of $R$,
it is also routine to show that
$R$ can convince $W$ with certainty
if ${b=1}$ is chosen by $W$ at Step~\ref{final verification step},
since ${P_r V_r \ket{\psi_r} = \ket{\psi_{r+1}}}$
for any $r$ chosen from ${\bigl\{1, \ldots, \frac{m}{2} \bigr\}}$.
Hence, $W$ accepts every input ${x \in A_{\yes}}$
with probability at least ${1 - \frac{\varepsilon}{2}}$.

Next, for the soundness, suppose that the input $x$ is in $A_{\no}$.

Let $R'$ be any three-message quantum prover
for the constructed proof system.
Let ${\rho_j \in \Density(\calV \otimes \calM)}$
be the reduced state in ${(\sfV_j, \sfM_j)}$ of the entire system state
just after the first transformation of $R'$,
for each ${1 \leq j \leq \frac{m}{2} + 1}$.

Consider the case
in which $W$ chooses $r$ from ${\bigl\{1, \ldots, \frac{m}{2} \bigr\}}$
in Step~\ref{verifier's question}
and also chooses ${b=1}$ at Step~\ref{final verification step}.
Then the probability that $R'$ can convince $W$
in this case cannot be larger than
${
  \frac{1}{2}
  +
  \frac{1}{2}
  F(\tr_{\calM} V_r \rho_r \conjugate{V_r}, \tr_{\calM} \rho_{r+1})
}$
by an argument similar to that in the proof of Theorem~4~in~Ref.~\cite{KitWat00STOC}.
Hence, the probability that $R'$ can convince $W$
when ${b=1}$ is chosen at Step~\ref{final verification step}
is at most
${
  \frac{1}{2}
  +
  \frac{1}{m}
  \sum_{j=1}^{\frac{m}{2}}
    F(\tr_{\calM} V_j \rho_j \conjugate{V_j}, \tr_{\calM} \rho_{j+1})
}$.

Now, if
${
  \tr \Pi_{\acc}
      V_{\frac{m}{2}+1}
      \rho_{\frac{m}{2} + 1}
      \conjugate{V_{\frac{m}{2}+1}}
  \geq
  1 - \frac{\delta}{4}
}$,
Lemma~\ref{Lemma: sum of sqrt of fidelity}
implies that
\[
\begin{split}
\hspace{1cm}&\hspace{-1cm}
\sum_{j=1}^{\frac{m}{2}}
  F(\tr_{\calM} V_j \rho_j \conjugate{V_j}, \tr_{\calM} \rho_{j+1})
\\
&
\leq
\frac{m}{2}
-
\frac{1}{m}
\left(
\sqrt{1 - \frac{\delta}{4}} - \sqrt{1 - \delta}
\right)^2
\leq
\frac{m}{2}
-
\frac{1}{m}
\left[
  \left( 1 - \frac{\delta}{4} \right)
  -
  \left( 1 - \frac{\delta}{2} \right)
\right]^2
=
\frac{m}{2} - \frac{\delta^2}{16m},
\end{split}
\]
and thus,
the probability that $R'$ can convince $W$
when ${b=1}$ is chosen
is at most
${
  \frac{1}{2}
  +
  \frac{1}{m}
  \left( \frac{m}{2} - \frac{\delta^2}{16m} \right)
  =
  1 - \frac{\delta^2}{16m^2}
}$.

On the other hand, if
${
  \tr \Pi_{\acc}
      V_{\frac{m}{2}+1}
      \rho_{\frac{m}{2} + 1}
      \conjugate{V_{\frac{m}{2}+1}}
  \leq
  1 - \frac{\delta}{4}
}$,
it is obvious that
$R'$ can convince $W$ with probability at most
${1 - \frac{\delta}{4} \leq 1 - \frac{\delta^2}{16m^2}}$
if ${b=0}$ is chosen by $W$ at Step~\ref{final verification step},
since the qubits in $\sfV_{\frac{m}{2}+1}$ and $\sfM_{\frac{m}{2}+1}$
are never touched by the prover after Step~\ref{first message}.

Hence the probability that $R'$ can convince $W$
for every input ${x \in A_{\no}}$
is at most ${1 - \frac{\delta^2}{32m^2}}$.
Taking it into account
that ${m(n)}$ may be odd for some ${n \in \Nonnegative}$,
we have the bound of ${1 - \frac{\delta^2}{32(m+1)^2}}$.

Finally, the perfect zero-knowledge property against $W$ is almost
straightforward.

Let $S_V$ be the simulator for the original $m$-message system
such that,
if $x$ is in $A_{\yes}$,
the states ${S_V(x,j)}$ and ${\view_{V,P}(x,j)}$
are identical for each ${1 \leq j \leq \frac{m}{2}}$.

The simulator $T_W$ for the constructed three-message system
behaves as follows.
For convenience,
let $\sfR$ be the quantum register
that is used to store the classical information $r$ chosen by $W$,
and let ${S_V(x,0) = \ketbra{0_{\calV \otimes \calM}}}$.

To simulate the state just after the first transformation of the prover $R$,
$T_W$ prepares the state ${S_V(x,j-1)}$ in ${(\sfV_j, \sfM_j)}$,
for each ${2 \leq j \leq \frac{m}{2}+1}$,
and outputs the state in 
${(\sfV_2, \sfM_2, \ldots, \sfV_{\frac{m}{2}+1}, \sfM_{\frac{m}{2}+1})}$
as ${T_W(x,1)}$.

To simulate the state just after the second transformation of the prover $R$,
$T_W$ first chooses ${r \in \bigl\{1, \ldots, \frac{m}{2} \bigr\}}$
uniformly at random, and sets the content of $\sfR$ to $r$.
Next $T_W$ prepares the state ${S_V(x,j-1)}$ in ${(\sfV_j, \sfM_j)}$,
for each ${1 \leq j \leq r-1}$ and ${r+1 \leq j \leq \frac{m}{2}+1}$,
and prepares the state ${S_V(x,r)}$ in ${(\sfV_r, \sfM_r)}$.
$T_W$ then prepares the state $\ket{\Phi^+}$ in ${(\sfX, \sfY)}$,
and performs a controlled-swap between
${(\sfV_r, \sfM_r)}$ and ${(\sfV_{r+1}, \sfM_{r+1})}$
using the qubit in $\sfX$ as the control.
Now $T_W$ outputs the state in
${(\sfR, \sfX, \sfY, \sfV_1, \sfM_1, \ldots, \sfV_{\frac{m}{2}+1}, \sfM_{\frac{m}{2}+1})}$
as ${T_W(x,2)}$.

It is obvious that the ensemble ${\{ T_W(x,j) \}}$
is polynomial-time preparable.

Suppose that $x$ is in $A_{\yes}$.

That ${T_W(x,1) = \view_{W,R}(x,1)}$ is obvious
from the fact that ${S_V(x,j) = \view_{V,P}(x,j)}$
for ${1 \leq j \leq \frac{m}{2}}$.

To show that ${T_W(x,2) = \view_{W,R}(x,2)}$,
let ${\view_{V,P}(x,0) = S_V(x,0) = \ketbra{0_{\calV \otimes \calM}}}$,
for convenience.
Let $\sigma_r$ and $\xi_r$ be the quantum states in
${(\sfR, \sfX, \sfY, \sfV_1, \sfM_1, \ldots, \sfV_{\frac{m}{2}+1}, \sfM_{\frac{m}{2}+1})}$
such that
\[
\sigma_r
=
\ketbra{r}
\otimes
\ketbra{\Phi^+}
\otimes
S_V(x,0) \otimes \cdots \otimes S_V(x,r-2)
\otimes
S_V(x,r)
\otimes
S_V(x,r) \otimes \cdots \otimes S_V \left( x,\frac{m}{2} \right)
\]
and
\[
\begin{split}
\hspace{1cm}&\hspace{-1cm}
\xi_r
=
\ketbra{r}
\otimes
\ketbra{\Phi^+}
\\
&
\otimes
\view_{V,P}(x,0) \otimes \cdots \otimes \view_{V,P}(x,r-2)
\otimes
\view_{V,P}(x,r)
\otimes
\view_{V,P}(x,r) \otimes \cdots \otimes \view_{V,P} \left( x,\frac{m}{2} \right)
\end{split}
\]
for each ${1 \leq r \leq \frac{m}{2}}$.
Then, we have ${\sigma_r = \xi_r}$ for each ${1 \leq r \leq \frac{m}{2}}$,
since ${S_V(x,j) = \view_{V,P}(x,j)}$ for ${0 \leq j \leq \frac{m}{2}}$.
For each ${1 \leq r \leq \frac{m}{2}}$,
let $\sigma'_r$ and $\xi'_r$
be the quantum states
obtained by performing a controlled-swap between
${(\sfV_r, \sfM_r)}$ and ${(\sfV_{r+1}, \sfM_{r+1})}$
on $\sigma_r$ and $\xi_r$, respectively,
using the qubit in $\sfX$ as the control.
Obviously, ${\sigma'_r = \xi'_r}$ for each ${1 \leq r \leq \frac{m}{2}}$.
By definition,
${T_W(x,2) = \frac{2}{m} \sum_{r=1}^{\frac{m}{2}} \sigma'_r}$.
Furthermore,
${\view_{W,R}(x,2)}$
is exactly the state
${\frac{2}{m} \sum_{r=1}^{\frac{m}{2}} \xi'_r}$.
Now that ${T_W(x,2) = \view_{W,R}(x,2)}$
follows from the fact that
${\sigma'_r = \xi'_r}$ for each ${1 \leq r \leq \frac{m}{2}}$.

Hence the honest-verifier perfect zero-knowledge property against $W$ follows.
\end{proof}

Next we show that
the parallel repetition theorem
for three-message quantum interactive proofs
may be extended to the case of
three-message honest-verifier quantum perfect zero-knowledge proof systems.

\begin{lemma}
Let
$\function{c, s}{\Nonnegative}{[0,1]}$
be any functions such that ${c > s}$.
Then,
for any polynomially bounded function
$\function{k}{\Nonnegative}{\Natural}$,
${\HVQPZK(3, c, s) \subseteq \HVQPZK(3, c^k, s^k)}$.
More strongly,
let $\Pi$ be any three-message honest-verifier quantum perfect zero-knowledge proof system
for a problem ${A = \{A_{\yes}, A_{\no}\}}$
with completeness accepting probability at least ${c(n)}$
and soundness accepting probability at most ${s(n)}$
for every input of length $n$.
Consider another proof system $\Pi'$
such that, for every input of length $n$,
$\Pi'$ carries out
${k(n)}$ attempts of $\Pi$ in parallel
and accepts
iff all the ${k(n)}$ attempts result in acceptance in $\Pi$.
Then $\Pi'$ is a three-message honest-verifier quantum perfect zero-knowledge proof system
for $A$
with completeness accepting probability at least ${c(n)^{k(n)}}$
and soundness accepting probability at most ${s(n)^{k(n)}}$
for every input of length $n$.
\label{Lemma: parallel repetition of three-message HVQPZK}
\end{lemma}

\begin{proof}
The completeness and soundness conditions follow
from the proof of Theorem~6 in Ref.~\cite{KitWat00STOC}.
The honest-verifier perfect zero-knowledge property is trivial.
Let $V$ be the honest quantum verifier
in the original three-message system $\Pi$
and let $S_V$ be the corresponding simulator 
such that,
if $x$ is in $A_{\yes}$,
the state ${S_V(x,j)}$ perfectly simulates $V$'s view
after the $j$th transformation of the honest quantum prover,
for each ${1 \leq j \leq 2}$.
Let $W$ be the honest quantum verifier
in the constructed three-message system $\Pi'$.
For every $x$ and for each ${1 \leq j \leq 2}$,
the simulator $T_W$ for $\Pi'$ just outputs
${T_W(x,j) = S_V(x,j)^{\otimes k(\abs{x})}}$.
Now the honest-verifier perfect zero-knowledge property is obvious.
\end{proof}

From Lemmas~\ref{Lemma: reducing number of messages of HVQPZK to three -- modified KW}~and~\ref{Lemma: parallel repetition of three-message HVQPZK},
it is immediate to show the following lemma.

\begin{lemma}
For any polynomially bounded function
$\function{p}{\Nonnegative}{\Natural}$,
${\HVQPZK \subseteq \HVQPZK(3, 1 - 2^{-p}, 2^{-p})}$.
\label{Lemma: parallelizing HVQPZK to three messages}
\end{lemma}

\begin{proof}
By sequential repetition,
we can show that,
for any polynomially bounded function $\function{m}{\Nonnegative}{\Natural}$,
for any functions $\function{c, s}{\Nonnegative}{[0,1]}$
that satisfy
${c - s \geq \frac{1}{q}}$
for some polynomially bounded function
$\function{q}{\Nonnegative}{\Natural}$,
and for any polynomially bounded function
$\function{p}{\Nonnegative}{\Natural}$,
there exists a polynomially bounded function
$\function{m'}{\Nonnegative}{\Natural}$
such that
${
\HVQPZK(m, c, s)
\subseteq
\HVQPZK(m', 1 - 2^{-p^2}, 2^{-p^2})
}$.
Now Lemma~\ref{Lemma: reducing number of messages of HVQPZK to three -- modified KW}
implies that
${
\HVQPZK(m', 1 - 2^{-p^2}, 2^{-p^2})
\subseteq
\HVQPZK \left(3, 1 - 2^{-p^2-1}, 1 - \frac{(1 - 2^{-p^2})^2}{32(m'+1)^2} \right)
}$.
Finally, by parallel repetition for sufficiently many times
(say, for ${32 p(\abs{x}) (m'(\abs{x})+2)^2}$ times),
from Lemma~\ref{Lemma: parallel repetition of three-message HVQPZK},
we have that
${
\HVQPZK \left(3, 1 - 2^{-p^2-1}, 1 - \frac{(1 - 2^{-p^2})^2}{32(m'+1)^2} \right)\subseteq
\HVQPZK(3, 1 - 2^{-p}, 2^{-p})
}$,
which completes the proof.
\end{proof}


\subsection{Converting Honest-Verifier Quantum Perfect Zero-Knowledge Proofs to Public-Coin Systems}
\label{Subsection: public-coin HVQPZK}

Next we show that
any three-message honest-verifier quantum perfect zero-knowledge system
can be modified to a three-message public-coin one
in which the message from the verifier
consists of only one classical bit.
Marriott~and~Watrous~\cite{MarWat05CC}
showed such a claim in the case of usual quantum interactive proofs.
We show that their construction
preserves the honest-verifier perfect zero-knowledge property.

\begin{lemma}
Let $\function{\varepsilon, \delta}{\Nonnegative}{[0,1]}$
be any functions that satisfy
${\delta > 1 - (1 - \varepsilon)^2}$.
Then,
any problem having
a three-message honest-verifier quantum perfect zero-knowledge system
with completeness accepting probability at least ${1 - \varepsilon}$
and soundness accepting probability at most ${1 - \delta}$
has
a three-message public-coin
honest-verifier quantum perfect zero-knowledge system
with completeness accepting probability at least ${1 - \frac{\varepsilon}{2}}$
and soundness accepting probability at most ${\frac{1}{2} + \frac{\sqrt{1 - \delta}}{2}}$
in which the message from the verifier
consists of only one classical bit.
\label{Lemma: three-message public-coin HVQPZK systems}
\end{lemma}

\begin{proof}
The proof is essentially same as that of Theorem~5.4~in~Ref.~\cite{MarWat05CC}
except for the zero-knowledge property.

Let ${A = \{A_{\yes}, A_{\no}\}}$ be a problem
in ${\HVQPZK(3, 1 - \varepsilon, 1 - \delta)}$
and let $V$ be the corresponding three-message quantum verifier.
Let $\sfV$ be the quantum register
consisting of all the qubits in the private space of $V$,
and let $\sfM$ be that
consisting of all the qubits in the message channel between $V$ and the prover.
For every input $x$,
$V$ applies $V_1$ and $V_2$ on the qubits in ${(\sfV, \sfM)}$
for his first and second transformations, respectively.
We construct a protocol of
a three-message public-coin quantum verifier $W$.

For every input $x$,
at the first message the constructed verifier $W$ receives
the quantum register $\sfV$ from the prover.
$W$ expects that
the prover prepares the quantum register $\sfM$ in his private space
and the qubits in ${(\sfV, \sfM)}$ form the quantum state
the original verifier $V$ would possess
just after the second message
(i.e., just after the first transformation of $V$)
of the original protocol.

At the second message,
$W$ chooses ${b \in \Binary}$ uniformly at random
and sends $b$ to the prover.

If ${b=0}$, the prover is requested to send $\sfM$,
so that the qubits in ${(\sfV, \sfM)}$ form the quantum state
the original verifier $V$ would possess
just after the third message
(i.e., just after the second transformation of the prover)
of the original protocol.
Now $W$ applies $V_2$ to the qubits in ${(\sfV, \sfM)}$
and accepts if and only if
the content of ${(\sfV, \sfM)}$
corresponds to an accepting state of the original protocol.

On the other hand,
if ${b=1}$, the prover is requested to send $\sfM$
so that the qubits in ${(\sfV, \sfM)}$ form the quantum state
the original verifier $V$ would possess
just after the second message
(i.e., just after the first transformation of $V$)
of the original protocol.
Now $W$ applies $\conjugate{V}_1$ to the qubits in ${(\sfV, \sfM)}$
and accepts if and only if
all the qubits in $\sfV$ are in state $\ket{0}$.

The precise description of the protocol of $W$ is found
in Figure~\ref{Figure: honest verifier's protocol in a three-message public-coin system}.

\begin{figure}
\begin{algorithm*}{Honest Verifier's Protocol in Three-Message Public-Coin System}
\begin{step}
\item
  Receive a quantum register $\sfV$ from the prover.
\item
  Choose ${b \in \{0, 1\}}$ uniformly at random.
  Send $b$ to the prover.
\item
  Receive a quantum register $\sfM$ from the prover.
  \begin{step}
  \item
    If ${b=0}$,
    apply $V_2$ to the qubits in ${(\sfV, \sfM)}$.
    Accept if the content of ${(\sfV, \sfM)}$
    corresponds to an accepting state of the original protocol,
    and reject otherwise.
  \item
    If ${b=1}$,
    apply $\conjugate{V}_1$ to the qubits in ${(\sfV, \sfM)}$.
    Accept if all the qubits in $\sfV$ are in state $\ket{0}$,
    and reject otherwise.
  \end{step}
\end{step}
\end{algorithm*}
\caption{Honest verifier's protocol in a three-message public-coin system.} 
\label{Figure: honest verifier's protocol in a three-message public-coin system}
\end{figure}

First suppose that the input $x$ is in $A_{\yes}$.

Let $P$ be the three-message honest quantum prover
for the original proof system,
and let $\sfP$ be the quantum register
consisting of all the qubits in the private space of $P$.
Let $\ket{\psi_2}$ be the quantum state in ${(\sfV, \sfM, \sfP)}$
just after the second message
(i.e., just after the first transformation of $V$)
of the original protocol
if $V$ communicates with $P$ on input $x$.

Let $R$ be the honest prover
in the constructed public-coin system.
In addition to the registers $\sfV$ and $\sfM$,
$R$ prepares the quantum register $\sfP$ in his private space.
At the first message of the constructed protocol,
$R$ first generates $\ket{\psi_2}$ in ${(\sfV, \sfM, \sfP)}$
and then sends $\sfV$ to $W$.

At the third message of the constructed protocol,
if ${b=0}$,
$R$ first applies $P_2$ to the qubits in ${(\sfM, \sfP)}$,
and then sends $\sfM$ to $W$,
where $P_2$ is the second transformation of the original prover $P$
on input $x$ in the original protocol,
while if ${b=1}$,
$R$ does nothing and just sends $\sfM$ to $W$.

It is obvious that
$R$ can convince $W$ with probability at least ${1 - \varepsilon}$
if ${b=0}$,
and with certainty if ${b=1}$.
Hence, $W$ accepts every input ${x \in A_{\yes}}$
with probability at least ${1 - \frac{\varepsilon}{2}}$.

The soundness property for the case the input $x$ is in $A_{\no}$
follows with exactly the same argument
as in the proof of Theorem~5.4~in~Ref.~\cite{MarWat05CC}.

Finally, the perfect zero-knowledge property against $W$ is almost
straightforward.

Let $S_V$ be the simulator for $V$ in the original system
such that,
if $x$ is in $A_{\yes}$,
the states ${S_V(x,j)}$ and ${\view_{V,P}(x,j)}$
are identical for each ${1 \leq j \leq 2}$.
Let $\calM$ be the Hilbert space
corresponding to the quantum register $\sfM$.
The simulator $T_W$ for the constructed public-coin system
behaves as follows.
For convenience,
let $\sfR$ be the single-qubit register
that is used to store the classical information
representing the outcome $b$ of a public coin flipped by $W$.

Let ${T_W(x,1)}$ and ${T_W(x,2)}$ be quantum states
in $\sfV$
and in ${(\sfR, \sfV, \sfM)}$, respectively,
defined by
\begin{align*}
T_W(x,1)
&=
\tr_{\calM} V_1 S_V(x,1) \conjugate{V_1},
\\
T_W(x,2)
&=
\frac{1}{2}
\bigl[
  \ketbra{0} \otimes S_V(x,2)
  +
  \ketbra{1} \otimes \bigl( V_1 S_V(x,1) \conjugate{V_1} \bigr)
\bigr].
\end{align*}
It is obvious that the ensemble ${\{ T_W(x,j) \}}$
is polynomial-time preparable.

Suppose that $x$ is in $A_{\yes}$.
It is obvious that
${T_W(x,1) = \view_{W,R}(x,1)}$,
since ${T_W(x,1) = \tr_{\calM} V_1 S_V(x,1) \conjugate{V_1}}$,
${\view(W,R)_1 = \tr_{\calM} V_1 \view_{V,P}(x,1) \conjugate{V_1}}$,
and ${S_V(x,1) = \view_{V,P}(x,1)}$.
The fact ${T_W(x,2) = \view_{W,R}(x,2)}$
follows from the properties
${
  \view_{W,R}(x,2)
  =
  \frac{1}{2}
  \bigl[
    \ketbra{0} \otimes \view_{V,P}(x,2)
    +
    \ketbra{1} \otimes \bigl( V_1 \view_{V,P}(x,1) \conjugate{V_1} \bigr)
  \bigr]
}$,
${S_V(x,1) = \view_{V,P}(x,1)}$,
and
${S_V(x,2) = \view_{V,P}(x,2)}$.

Hence the claim follows.
\end{proof}


\subsection{$\boldsymbol{\HVQPZK = \QPZK}$}
\label{Subsection: HVQPZK = QPZK}

First notice that
the quantum rewinding technique due to Watrous~\cite{Wat06STOC}
perfectly works well for any three-message public-coin
honest-verifier quantum perfect zero-knowledge protocol
in which the message from the verifier
consists of only one classical bit.
That is, we can show the following lemma.

\begin{lemma}
Any three-message public-coin 
honest-verifier quantum perfect zero-knowledge system
such that the message from the verifier
consists of only one classical bit
is perfect zero-knowledge against
any polynomial-time quantum verifier.
\label{Lemma: perfect zero-knowledge against any verifier}
\end{lemma}

\begin{proof}
Let ${A = \{A_{\yes}, A_{\no}\}}$ be a problem
having a three-message public-coin 
honest-verifier quantum perfect zero-knowledge system
such that the message from the verifier
consists of only one classical bit.
Let $V$ and $P$ be the corresponding
three-message public-coin honest quantum verifier
and three-message honest quantum prover,
respectively.
Let $\sfM$ and $\sfN$ be the quantum registers
consisting of all the qubits sent to $V$ at the first message
and of those at the third message, respectively,
and let $\sfR$ and $\sfS$ be the single-qubit registers
that are used to store the classical information
representing the outcome $b$ of a public coin flipped by $V$,
where $\sfR$ is inside the private space of $V$
and $\sfS$ is sent to $P$.

Let $S_V$ be the simulator for $V$
such that,
if $x$ is in $A_{\yes}$,
the states ${S_V(x,1)}$ and ${\view_{V,P}(x,1)}$
consisting of qubits in $\sfM$ are identical
and the states ${S_V(x,2)}$ and ${\view_{V,P}(x,2)}$
consisting of qubits in ${(\sfM, \sfN, \sfR)}$ are also identical.

Consider a generating circuit $Q$ of the quantum state ${S_V(x,2)}$.
Without loss of generality,
it is assumed that $Q$ acts over the qubits in ${(\sfM, \sfN, \sfR, \sfA)}$,
where $\sfA$ is the quantum register consisting of $q_{\calA}$ qubits
for some polynomially bounded function
$\function{q_{\calA}}{\Nonnegative}{\Natural}$.

For any polynomial-time quantum verifier $W$
and any auxiliary quantum state $\rho$ for $W$
stored in the quantum register $\sfX$ inside the private space of $W$,
we construct an efficiently implementable admissible mapping $\Phi$
that corresponds to a simulator $T_W$ for $W$.
Without loss of generality
it is assumed that the message from $W$ consists of a single \emph{classical} bit,
since the honest prover can easily enforce this constraint
by measuring the message from the verifier before responding to it.
Let $\sfW$ be the quantum register consisting of all the qubits
in the private space of $W$ except for those in $\sfX$ and $\sfM$
after having sent the second message.
We consider the procedure described
in Figure~\ref{Figure: simulator for general verifier},
which is the implementation of $\Phi$.

\begin{figure}
\begin{algorithm*}{Simulator for General Verifier $\boldsymbol{W}$}
\begin{step}
\item
  Store the auxiliary quantum state $\rho$ in the quantum register $\sfX$.
  Prepare the quantum registers
  $\sfS$, $\sfW$, $\sfM$, $\sfN$, $\sfR$, and $\sfA$,
  and further prepare a single qubit quantum register $\sfF$.
  Initialize all the qubits in
  $\sfF$, $\sfS$, $\sfW$, $\sfM$, $\sfN$, $\sfR$, and $\sfA$
  in state $\ket{0}$.
\item
  Apply the generating circuit $Q$ of the quantum state ${S_V(x,2)}$
  to the qubits in ${(\sfM, \sfN, \sfR, \sfA)}$.%
  \label{preparation step}
\item
  Apply $W_1$ to the qubits in ${(\sfS, \sfW, \sfX, \sfM)}$,
  where $W_1$ is the first transformation of the simulated verifier $W$.
\item
  Compute the exclusive-or of the contents of $\sfR$ and $\sfS$
  and write the result in $\sfF$.
\item
  Measure the qubit in $\sfF$ in the ${\{ \ket{0}, \ket{1} \}}$ basis.
  If this results in $\ket{0}$,
  output the qubits in ${(\sfW, \sfX, \sfM, \sfN, \sfR)}$,
  otherwise apply $\conjugate{W_1}$
  to the qubits in ${(\sfS, \sfW, \sfX, \sfM)}$
  and then apply $\conjugate{Q}$
  to the qubits in ${(\sfM, \sfN, \sfR, \sfA)}$.%
  \label{measuring step}
\item
  Apply the phase-flip
  if all the qubits in $\sfF$, $\sfS$, $\sfW$, $\sfM$, $\sfN$, $\sfR$, and $\sfA$
  are in state $\ket{0}$,
  apply $Q$ to the qubits in ${(\sfM, \sfN, \sfR, \sfA)}$,
  and apply $W_1$ to the qubits in ${(\sfS, \sfW, \sfX, \sfM)}$.
  Output the qubits in ${(\sfW, \sfX, \sfM, \sfN, \sfR)}$.%
  \label{the end of rewinding}
\end{step}
\end{algorithm*}
\caption{Simulator for a general verifier $W$.}
\label{Figure: simulator for general verifier}
\end{figure}

Suppose that the input $x$ is in $A_{\yes}$.

Since the state ${\view_{V,P}(x,2)}$ can be written of the form
${
  \view_{V,P}(x,2)
  =
  \frac{1}{2} (\sigma_0 \otimes \ketbra{0} + \sigma_1 \otimes \ketbra{1})
}$
for some quantum states $\sigma_0$ and $\sigma_1$ in ${(\sfM, \sfN)}$,
the state ${S_V(x,2)}$ must also be of the form
${
  S_V(x,2)
  =
  \frac{1}{2} (\sigma_0 \otimes \ketbra{0} + \sigma_1 \otimes \ketbra{1})
}$
from the honest-verifier perfect zero-knowledge property.
Therefore, the probability of obtaining $\ket{0}$
as the measurement result in Step~\ref{measuring step}
is exactly equal to $\frac{1}{2}$
regardless of the auxiliary quantum state $\rho$,
because
${\tr_{\calN} \sigma_0 = \tr_{\calN} \sigma_1}$
holds from the honest-verifier perfect zero-knowledge property of the protocol,
where $\calN$ is the Hilbert space corresponding to $\sfN$
(recall that when communicating with the honest verifier $V$,
the qubits in $\sfM$ are never touched by $V$ until the final
transformation of $V$).

Let
${
\xi_i
=
\Pi_i W_1
(\ketbra{0_{\calS \otimes \calW}} \otimes \rho \otimes \sigma_i \otimes \ketbra{i})
\conjugate{W_1} \Pi_i
}$
be an unnormalized state in ${(\sfS, \sfW, \sfX, \sfM, \sfN, \sfR)}$
for each ${i \in \Binary}$,
where ${\Pi_i = \ketbra{i}}$ is the projection operator
over the qubit in $\sfS$,
and $\calS$ and $\calW$ are the Hilbert spaces
corresponding to $\sfS$ and $\sfW$, respectively.
Then, conditioned on the measurement result being $\ket{0}$
in Step~\ref{measuring step},
the output is the state
${\tr_{\calS} (\xi_0 + \xi_1)}$.

Noticing that $\tr_{\calS} \frac{\xi_i}{\tr \xi_i}$
is exactly the state the verifier $W$ would possess
after the third message
when the second message from $W$ is $i$
and that the probability of the second message from $W$
being $i$ is exactly equal to $\tr \xi_i$
for each ${i \in \Binary}$,
${
  \tr_{\calS} (\xi_0 + \xi_1)
  =
  \tr \xi_0 \cdot \tr_{\calS} \frac{\xi_0}{\tr \xi_0}
  +
  \tr \xi_1 \cdot \tr_{\calS} \frac{\xi_1}{\tr \xi_1}
}$
is exactly the state $W$ would possess
after the third message.
Thus, the quantum rewinding technique due to Watrous~\cite{Wat06STOC}
perfectly works well,
which is implemented in Steps~\ref{measuring step}~and~\ref{the end of rewinding}.

This ensures the perfect zero-knowledge property against $W$,
which completes the proof.
\end{proof}

From Lemma~\ref{Lemma: perfect zero-knowledge against any verifier},
it is immediate to show that ${\HVQPZK = \QPZK}$,
i.e., 
honest-verifier quantum perfect zero-knowledge
equals general quantum perfect zero-knowledge.

\begin{theorem}
${\HVQPZK = \QPZK}$.
\label{Theorem: HVQPZK = QPZK}
\end{theorem}

\begin{proof}
That ${\HVQPZK \supseteq \QPZK}$ is trivial
and we show that ${\HVQPZK \subseteq \QPZK}$.
Now Lemma~\ref{Lemma: perfect zero-knowledge against any verifier}
together with
Lemmas~\ref{Lemma: parallelizing HVQPZK to three messages}~and~\ref{Lemma: three-message public-coin HVQPZK systems}
implies that
${
  \HVQPZK
  \subseteq
  \QPZK \left(3, 1 - 2^{-p}, \frac{1}{2} + 2^{-\frac{p}{2}-1} \right)
}$
for any polynomially bounded function
$\function{p}{\Nonnegative}{\Natural}$.
Therefore, the fact that sequential repetition works well
for the protocols of quantum zero-knowledge proofs
establishes the statement.
\end{proof}

From the proof of Theorem~\ref{Theorem: HVQPZK = QPZK},
the following property also follows.

\begin{theorem}
Any problem in $\QPZK$
has a public-coin quantum perfect zero-knowledge proof system.
\label{Theorem: public-coin QPZK = QPZK}
\end{theorem}


\section{Computational Zero-Knowledge Case}
\label{Section: QZK}


\subsection{$\boldsymbol{\HVQZK = \QZK}$}
\label{Subsection: HVQZK = QZK}

With essentially same arguments as in the perfect zero-knowledge case,
we can show that honest-verifier quantum zero-knowledge
equals general quantum zero-knowledge
for the computational zero-knowledge case.


First, we show the following lemma,
which is the computational zero-knowledge version of
Lemma~\ref{Lemma: reducing number of messages of HVQPZK to three -- modified KW}.
The proof is exactly the same as the proof of
Lemma~\ref{Lemma: reducing number of messages of HVQPZK to three -- modified KW}
except for the zero-knowledge property
and the honest-verifier computational zero-knowledge property
can be proved by fairly straightforward hybrid arguments.

\begin{lemma}
Let $\function{m}{\Nonnegative}{\Natural}$
be a polynomially bounded function
and let
$\function{\varepsilon, \delta}{\Nonnegative}{[0,1]}$
be any functions
such that ${m \geq 4}$ and ${\varepsilon < \frac{\delta^2}{16(m+1)^2}}$.
Then,
${
  \HVQZK(m, 1 - \varepsilon, 1 - \delta)
  \subseteq
  \HVQZK \left( 3, 1 - \frac{\varepsilon}{2}, 1 - \frac{\delta^2}{32(m+1)^2} \right)
}$.
\label{Lemma: reducing number of messages of HVQZK to three -- modified KW}
\end{lemma}

Alternatively, we may show
the computational zero-knowledge version of
Theorem~4~in~Ref.~\cite{KitWat00STOC}.

Next we show that the parallel repetition theorem
for three-message quantum interactive proofs
may be extended to the case of
three-message honest-verifier quantum computational zero-knowledge proof systems,
which is the the computational zero-knowledge version of
Lemma~\ref{Lemma: parallel repetition of three-message HVQPZK}.
Again the proof is exactly the same as the proof of
Lemma~\ref{Lemma: parallel repetition of three-message HVQPZK}
except for the zero-knowledge property
and the honest-verifier computational zero-knowledge property
can be proved by fairly straightforward hybrid arguments.

\begin{lemma}
Let
$\function{c, s}{\Nonnegative}{[0,1]}$
be any functions such that ${c > s}$.
Then,
for any polynomially bounded function
$\function{k}{\Nonnegative}{\Natural}$,
${\HVQZK(3, c, s) \subseteq \HVQZK(3, c^k, s^k)}$.
More strongly,
let $\Pi$ be any three-message honest-verifier quantum computational zero-knowledge proof system
for a problem ${A = \{A_{\yes}, A_{\no}\}}$
with completeness accepting probability at least ${c(n)}$
and soundness accepting probability at most ${s(n)}$
for every input of length $n$.
Consider another proof system $\Pi'$
such that, for every input of length $n$,
$\Pi'$ carries out
${k(n)}$ attempts of $\Pi$ in parallel
and accepts
iff all the ${k(n)}$ attempts result in acceptance in $\Pi$.
Then $\Pi'$ is a three-message honest-verifier quantum computational zero-knowledge proof system
for $A$
with completeness accepting probability at least ${c(n)^{k(n)}}$
and soundness accepting probability at most ${s(n)^{k(n)}}$
for every input of length $n$.
\label{Lemma: parallel repetition of three-message HVQZK}
\end{lemma}


Now Lemma~\ref{Lemma: parallelizing HVQZK to three messages} below
follows from the essentially same argument as in the proof of
Lemma~\ref{Lemma: parallelizing HVQPZK to three messages},
using Lemmas~\ref{Lemma: reducing number of messages of HVQZK to three -- modified KW}~and~\ref{Lemma: parallel repetition of three-message HVQZK}.

\begin{lemma}
For any polynomially bounded function
$\function{p}{\Nonnegative}{\Natural}$,
${\HVQZK \subseteq \HVQZK(3, 1 - 2^{-p}, 2^{-p})}$.
\label{Lemma: parallelizing HVQZK to three messages}
\end{lemma}

We can also show the following lemma,
which is the computational zero-knowledge version of
Lemma~\ref{Lemma: three-message public-coin HVQPZK systems}.

\begin{lemma}
Let $\function{\varepsilon, \delta}{\Nonnegative}{[0,1]}$
be any functions that satisfy
${\delta > 1 - (1 - \varepsilon)^2}$.
Then,
any problem having
a three-message honest-verifier quantum computational zero-knowledge system
with completeness accepting probability at least ${1 - \varepsilon}$
and soundness accepting probability at most ${1 - \delta}$
has
a three-message public-coin
honest-verifier quantum computational zero-knowledge system
with completeness accepting probability at least ${1 - \frac{\varepsilon}{2}}$
and soundness accepting probability at most ${\frac{1}{2} + \frac{\sqrt{1 - \delta}}{2}}$
in which the message from the verifier
consists of only one classical bit.
\label{Lemma: three-message public-coin HVQZK systems}
\end{lemma}

\begin{proof}
We use the same protocol construction as in the proof of
Lemma~\ref{Lemma: three-message public-coin HVQPZK systems}
and we only show the zero-knowledge property.
In what follows, we use the same notations as in the proof of
Lemma~\ref{Lemma: three-message public-coin HVQPZK systems}.

Let $S_V$ be the simulator for the original system
such that,
if $x$ is in $A_{\yes}$,
the states ${S_V(x,j)}$ and ${\view_{V,P}(x,j)}$
are computationally indistinguishable for each ${1 \leq j \leq 2}$.
Let $\calM$ be the Hilbert space
corresponding to the quantum register $\sfM$.
As in the proof of
Lemma~\ref{Lemma: three-message public-coin HVQPZK systems},
the simulator $T_W$ for the constructed public-coin system
behaves as follows.
For convenience,
as in the proof of
Lemma~\ref{Lemma: three-message public-coin HVQPZK systems},
let $\sfR$ be the single-qubit register
that is used to store the classical information
representing the outcome $b$ of a public coin flipped by $W$.

Let ${T_W(x,1)}$ and ${T_W(x,2)}$ be quantum states
in $\sfV$
and in ${(\sfR, \sfV, \sfM)}$, respectively,
defined by
\begin{align*}
T_W(x,1)
&=
\tr_{\calM} V_1 S_V(x,1) \conjugate{V_1},
\\
T_W(x,2)
&=
\frac{1}{2}
\bigl[
  \ketbra{0} \otimes S_V(x,2)
  +
  \ketbra{1} \otimes \bigl( V_1 S_V(x,1) \conjugate{V_1} \bigr)
\bigr].
\end{align*}
It is obvious that the ensemble ${\{ T_W(x,j) \}}$
is polynomial-time preparable.

Suppose that $x$ is in $A_{\yes}$.
The computational indistinguishability between
${T_W(x,1)}$ and ${\view_{W,R}(x,1)}$
is obvious
since ${T_W(x,1) = \tr_{\calM} V_1 S_V(x,1) \conjugate{V_1}}$,
${\view(W,R)_1 = \tr_{\calM} V_1 \view_{V,P}(x,1) \conjugate{V_1}}$,
and ${S_V(x,1)}$ and ${\view_{V,P}(x,1)}$
are computational indistinguishable.
The computational indistinguishability between
${T_W(x,2)}$ and ${\view_{W,R}(x,2)}$
follows from the properties
${
  \view_{W,R}(x,2)
  =
  \frac{1}{2}
  \bigl[
    \ketbra{0} \otimes \view_{V,P}(x,2)
    +
    \ketbra{1} \otimes \bigl( V_1 \view_{V,P}(x,1) \conjugate{V_1} \bigr)
  \bigr]
}$,
the computational indistinguishability
between ${S_V(x,1)}$ and ${\view_{V,P}(x,1)}$,
and that between ${S_V(x,2)}$ and ${\view_{V,P}(x,2)}$.

Now the lemma follows.
\end{proof}

Now applying the quantum rewinding technique due to Watrous~\cite{Wat06STOC},
we show the computational zero-knowledge version of
Lemma~\ref{Lemma: perfect zero-knowledge against any verifier},
that any three-message public-coin 
honest-verifier quantum computational zero-knowledge system
such that the message from the verifier
consists of only one classical bit
is computational zero-knowledge against
any dishonest quantum verifier.

\begin{lemma}
Any three-message public-coin 
honest-verifier quantum computational zero-knowledge system
such that the message from the verifier
consists of only one classical bit
is computational zero-knowledge against
any polynomial-time quantum verifier.
\label{Lemma: computational zero-knowledge against any verifier}
\end{lemma}

\begin{proof}
We use the same construction of the simulator as in the proof of
Lemma~\ref{Lemma: perfect zero-knowledge against any verifier}.
In what follows, we use the same notations as in the proof of
Lemma~\ref{Lemma: perfect zero-knowledge against any verifier}.

Let $S_V$ be the simulator for $V$
such that,
if $x$ is in $A_{\yes}$,
the states ${S_V(x,1)}$ and ${\view_{V,P}(x,1)}$
consisting of qubits in $\sfM$
are computationally indistinguishable
and the states ${S_V(x,2)}$ and ${\view_{V,P}(x,2)}$
consisting of qubits in ${(\sfM, \sfN, \sfR)}$
are also computationally indistinguishable,
and consider the simulator construction
in Figure~\ref{Figure: simulator for general verifier}
in the proof of Lemma~\ref{Lemma: perfect zero-knowledge against any verifier}.

Suppose that the input $x$ is in $A_{\yes}$.

We shall show that
(i) the gap between $\frac{1}{2}$ and
the probability of obtaining $\ket{0}$
as the measurement result in Step~\ref{measuring step}
must be negligible
regardless of the auxiliary quantum state $\rho$,
and (ii) the output state in Step~\ref{measuring step} in the construction
conditioned on the measurement result being $\ket{0}$
must be computationally indistinguishable
from the state $W$ would possess
after the third message.
With these two properties,
the quantum rewinding technique due to Watrous~\cite{Wat06STOC}
works well,
by using the amplification lemma for the case with negligible perturbations,
which is also due to Watrous~\cite{Wat06STOC}.
This ensures the computational zero-knowledge property against $W$.

For the generating circuit $Q'$ of the quantum state ${\view_{V,P}(x,2)}$
(for example, the unitary circuit $P_1$
that corresponds to the first transformation of the honest prover $P$
realizes $Q'$),
consider the ``ideal'' construction of the simulator
such that $Q'$ is applied instead of $Q$
in Step~\ref{preparation step} of the ``real'' simulator construction.

We first show the property (i).

Since the state ${\view_{V,P}(x,2)}$ can be written of the form
${
  \view_{V,P}(x,2)
  =
  \frac{1}{2} (\sigma_0 \otimes \ketbra{0} + \sigma_1 \otimes \ketbra{1})
}$
for some quantum states $\sigma_0$ and $\sigma_1$ in ${(\sfM, \sfN)}$,
the probability of obtaining $\ket{0}$
as the measurement result in Step~\ref{measuring step}
in the ``ideal'' construction
is exactly equal to $\frac{1}{2}$
regardless of the auxiliary quantum state $\rho$,
because
${\tr_{\calN} \sigma_0 = \tr_{\calN} \sigma_1}$
necessarily holds in this case,
where $\calN$ is the Hilbert space corresponding to $\sfN$.

Now, from the honest-verifier computational zero-knowledge property,
the states ${S_V(x,2)}$ and ${\view_{V,P}(x,2)}$
in ${(\sfM, \sfN, \sfR)}$
are computationally indistinguishable.
Since the circuit implementing $W_1$ is of size polynomial
with respect to $\abs{x}$,
it follows that
the gap between $\frac{1}{2}$ and
the probability of obtaining $\ket{0}$
as the measurement result in Step~\ref{measuring step}
in the ``real'' construction
must be negligible
regardless of the auxiliary quantum state $\rho$,
which proves the property (i).

Now we show the property (ii).

Let
${
\xi_i
=
\Pi_i W_1
(\ketbra{0_{\calS \otimes \calW}} \otimes \rho \otimes \sigma_i \otimes \ketbra{i})
\conjugate{W_1} \Pi_i
}$
be an unnormalized state in ${(\sfS, \sfW, \sfX, \sfM, \sfN, \sfR)}$
for each ${i \in \Binary}$,
where ${\Pi_i = \ketbra{i}}$ is the projection operator
over the qubits in $\sfS$,
and $\calS$ and $\calW$ are the Hilbert spaces
corresponding to $\sfS$ and $\sfW$, respectively.
Then, in the ``ideal'' construction,
conditioned on the measurement result being $\ket{0}$
in Step~\ref{measuring step},
the output is the state
${\tr_{\calS} (\xi_0 + \xi_1)}$.

Noticing that $\tr_{\calS} \frac{\xi_i}{\tr \xi_i}$
is exactly the state the verifier $W$ would possess
after the third message
when the second message from $W$ is $i$
and that the probability of the second message from $W$
being $i$ is exactly equal to $\tr \xi_i$
for each ${i \in \Binary}$,
${
  \tr_{\calS} (\xi_0 + \xi_1)
  =
  \tr \xi_0 \cdot \tr_{\calS} \frac{\xi_0}{\tr \xi_0}
  +
  \tr \xi_1 \cdot \tr_{\calS} \frac{\xi_1}{\tr \xi_1}
}$
is exactly the state $W$ would possess
after the third message.

Towards a contradiction,
suppose that the output state in Step~\ref{measuring step} in the ``real'' construction
conditioned on the measurement result being $\ket{0}$
is computationally distinguishable
from ${\tr_{\calS} (\xi_0 + \xi_1)}$,
which is the state $W$ would possess
after the third message.
Let $D$ be the corresponding distinguisher
that uses the auxiliary quantum state $\rho'$.
We construct a distinguisher $D'$
for ${S_V(x,2)}$ and ${\view_{V,P}(x,2)}$ from $D$.

On input quantum state $\xi$
that is either ${S_V(x,2)}$ or ${\view_{V,P}(x,2)}$,
$D'$ uses the auxiliary quantum state ${\rho \otimes \rho'}$,
where $\rho$ is the auxiliary quantum state the verifier $W$ would use.
$D'$ prepares the quantum registers
$\sfS$, $\sfW$, $\sfM$, $\sfN$, $\sfR$
and another quantum register $\sfY$.
$D'$ stores $\rho$ in the register $\sfX$,
$\xi$ in the register ${(\sfM, \sfN, \sfR)}$,
and $\rho'$ in $\sfY$.
All the qubits in $\sfS$ and $\sfW$ are initialized in state $\ket{0}$.
Now $D'$ applies $W_1$ to the qubits in ${(\sfS, \sfW, \sfX, \sfM)}$,
and then applies $D$ to the qubits in
${(\sfW, \sfX, \sfM, \sfN, \sfR, \sfY)}$.

It is obvious from this construction
that $D'$ with the auxiliary quantum state ${\rho \otimes \rho'}$
forms a distinguisher
for ${S_V(x,2)}$ and ${\view_{V,P}(x,2)}$
if $D$ with the auxiliary quantum state $\rho'$
forms a distinguisher
for the output state in Step~\ref{measuring step}
in the ``real'' simulator construction
conditioned on the measurement result being $\ket{0}$
and the state ${\tr_{\calS} (\xi_0 + \xi_1)}$.
This contradicts the computational indistinguishability
between ${S_V(x,2)}$ and ${\view_{V,P}(x,2)}$,
and thus the property (ii) follows.
\end{proof}

From Lemmas~\ref{Lemma: parallelizing HVQZK to three messages},~\ref{Lemma: three-message public-coin HVQZK systems},~and~\ref{Lemma: computational zero-knowledge against any verifier},
it is easy to show that
honest-verifier quantum computational zero-knowledge
equals general quantum computational zero-knowledge.
The proof is essentially same as the proof of Theorem~\ref{Theorem: HVQPZK = QPZK},
and thus, the property that
public-coin quantum computational zero-knowledge
equals general quantum computational zero-knowledge
also follows.

\begin{theorem}
${\HVQZK = \QZK}$.
\label{Theorem: HVQZK = QZK}
\end{theorem}

\begin{theorem}
Any problem in $\QZK$
has a public-coin quantum computational zero-knowledge proof system.
\label{Theorem: public-coin QZK = QZK}
\end{theorem}


\subsection{QZK with Perfect Completeness Equals General QZK}
\label{Subection: QZK with Perfect Completeness Equals General QZK}

In the computational zero-knowledge case,
we can show that
quantum computational zero-knowledge
with one-sided bounded error of perfect completeness
equals general quantum computational zero-knowledge.

The key idea is to show that
any \emph{honest-verifier} quantum computational zero-knowledge proof system
with two-sided bounded error
can be modified to that with one-sided bounded error of perfect completeness.
This can be proved in a similar manner as
in the proof of Theorem~2~of~Ref.~\cite{KitWat00STOC},
but requires more careful analyses
for showing the zero-knowledge property.

\begin{lemma}
Let $\function{m}{\Nonnegative}{\Natural}$
be a polynomially bounded function,
let
$\function{\varepsilon}{\Nonnegative}{[0,1]}$
be any negligible function
such that
there exists a polynomial-time uniformly generated
family ${\{Q_{1^n}\}}$ of quantum circuits
such that $Q_{1^n}$ exactly performs the unitary transformation
\[
U_{\varepsilon(n)}
=
\begin{pmatrix}
\sqrt{\varepsilon(n)} & \sqrt{1 - \varepsilon(n)}\\
\sqrt{1 - \varepsilon(n)} & -\sqrt{\varepsilon(n)}
\end{pmatrix},
\]
and let
$\function{\delta}{\Nonnegative}{[0,1]}$
be any function that satisfies
${\delta > \varepsilon}$.
Then,
${
  \HVQZK(m, 1 - \varepsilon, 1 - \delta)
  \subseteq
  \HVQZK(m+2, 1, 1 - (\delta - \varepsilon)^2)
}$.
\label{Lemma: HVQZK with perfect completeness}
\end{lemma}

\begin{proof}
The proof is similar to
the proof of Theorem~2~of~Ref.~\cite{KitWat00STOC},
but requires more careful analyses
for showing the zero-knowledge property.

Let ${A = \{A_{\yes}, A_{\no}\}}$ be a problem
in ${\HVQZK(m, 1 - \varepsilon, 1 - \delta)}$,
and let $V$ be
the corresponding
$m$-message honest quantum verifier.
Let $\sfV$ be the quantum register
consisting of all the qubits in the private space of $V$,
and let $\sfM$ be that
consisting of all the qubits
in the message channel between $V$ and the prover.
For every input $x$,
$V$ applies $V_j$ for his $j$th transformation
to the qubits in ${(\sfV, \sfM)}$,
for ${1 \leq j \leq \floorL{\frac{m}{2}} + 1}$.
We construct a protocol of
an ${(m+2)}$-message honest quantum verifier $W$.
For simplicity,
in what follows, it is assumed that $m$ is even
(the cases in which $m$ is odd can be proved in a similar manner).

For every input $x$,
the new verifier $W$ prepares
the quantum registers $\sfV$ and $\sfM$
and another single-qubit quantum register $\sfX$.
Let $\sfY$ be the single-qubit quantum register
consisting of the qubit in $\sfV$
that corresponds to the output qubit of the original verifier $V$.

Using first ${(m-1)}$ messages,
$W$ attempts to simulate the first ${(m-1)}$ messages
of the original $m$-message protocol,
by applying $V_j$ to the qubits in ${(\sfV, \sfM)}$
as his $j$th transformation,
for ${1 \leq j \leq \frac{m}{2}}$.

At the $m$th message, which is from the prover,
$W$ receives a single-qubit quantum register $\sfB$
in addition to $\sfM$.
$W$ then applies $V_{\frac{m}{2}+1}$ to the qubits in ${(\sfV, \sfM)}$,
and further performs the Toffoli transformation
over the qubits in ${(\sfB, \sfY, \sfX)}$,
using the qubit in $\sfX$ as the target.
Notice that the content of $\sfX$ is $1$
if and only if the content of $\sfB$ is $1$
and the state in ${(\sfV, \sfM)}$
is an accepting state of the original protocol.
Then $W$ sends the registers $\sfB$, $\sfV$, and $\sfM$ to the prover,
while keeping only $\sfX$ in his private.

At the last message of the protocol,
$W$ receives the qubit in $\sfB$ 
and verifies if
the qubits in ${(\sfX, \sfB)}$ form the state
${
  \ket{\phi}
  =
  \sqrt{\varepsilon} \ket{00} + \sqrt{1 - \varepsilon} \ket{11}
}$.

The precise description of the protocol of $W$ is described
in Figure~\ref{Figure: Honest Verifier's Protocol for Achieving Perfect Completeness}.

\begin{figure}
\begin{algorithm*}{Honest Verifier's Protocol for Achieving Perfect Completeness}
\begin{step}
\item
  Prepare quantum registers $\sfV$ and $\sfM$
  and a single-qubit quantum register $\sfX$.
  Let $\sfY$ be the single-qubit quantum register
  consisting of the qubit in $\sfV$
  that corresponds to the output qubit of the original verifier.
  Initialize all the qubits in $\sfV$, $\sfM$, and $\sfX$
  in state $\ket{0}$.
  Apply $V_1$ to the qubits in ${(\sfV, \sfM)}$,
  and send $\sfM$ to the prover.
\item
  For ${j = 2}$ to $\frac{m}{2}$, do the following:\\
  Receive $\sfM$ from the prover.
  Apply $V_j$ to the qubits in ${(\sfV, \sfM)}$,
  and send $\sfM$ to the prover.
\item
  Receive $\sfB$ and $\sfM$ from the prover.
  Apply $V_{\frac{m}{2}+1}$ to the qubits in ${(\sfV, \sfM)}$
  and perform the Toffoli transformation over the qubits in ${(\sfX, \sfY, \sfB)}$
  using the qubit in $\sfX$ as the target.
  Send $\sfV$, $\sfM$, and $\sfB$ to the prover.
\item
  Receive $\sfB$ from the prover.
  Perform a controlled-not over the qubits in ${(\sfX, \sfB)}$
  using the qubit in $\sfX$ as the control.
  Apply $\conjugate{U_{\varepsilon}}$ to the qubit in $\sfX$.
  Accept if the content of $\sfX$ is $0$,
  and reject otherwise.
\end{step}
\end{algorithm*}
\caption{Honest verifier's protocol for achieving perfect completeness}
\label{Figure: Honest Verifier's Protocol for Achieving Perfect Completeness}
\end{figure}

The soundness can be proved in almost the same way as
in the proof of Theorem~2~of~Ref.~\cite{KitWat00STOC}.
We show the completeness and the honest-verifier zero-knowledge properties.
We first describe how the honest quantum prover behaves
in the constructed ${(m+2)}$-message system.

Suppose that the input $x$ is in $A_{\yes}$.
Let $P$ be the $m$-message honest quantum prover
for the original proof system,
and suppose that ${(V, P)}$ accepts $x$
with probability exactly ${p_{\acc} \geq 1 - \varepsilon}$.
Let $\sfP$ be the quantum register
consisting of all the qubits in the private space of $P$.
Let $P_j$ be the $j$th transformation of $P$
on input $x$ in the original protocol,
for ${1 \leq j \leq \frac{m}{2}}$.

The ${(m+2)}$-message honest quantum prover $R$
for the constructed proof system
prepares the register $\sfP$
and another single-qubit quantum register $\sfB$
in his private space.
All the qubits in $\sfP$ and $\sfB$ are initially in state $\ket{0}$.

At the $j$th transformation of $R$,
for ${1 \leq j \leq \frac{m}{2} - 1}$,
after receiving the register $\sfM$ from $W$,
$R$ applies $P_j$ to the qubits in ${(\sfM, \sfP)}$
and sends $\sfM$ to $W$.

At the $\frac{m}{2}$th transformation of $R$,
after receiving the register $\sfM$ from $W$,
$R$ first applies $P_j$ to the qubits in ${(\sfM, \sfP)}$.
$R$ also generates the state
${
  \ket{b}
  =
  \sqrt{1 - \frac{1-\varepsilon}{p_{\acc}}} \ket{0}
  +
  \sqrt{\frac{1-\varepsilon}{p_{\acc}}} \ket{1}
}$
in the register $\sfB$,
and sends $\sfB$ and $\sfM$ to $W$.

Let $\ket{\psi_{m+1}}$ be the system state
in ${(\sfX, \sfV, \sfM, \sfB)}$
just after the ${(m+1)}$-st message of the constructed protocol,
when $W$ is communicating with $R$ on the input $x$.
Then $\ket{\psi_{m+1}}$ can be written as
${
  \ket{\psi_{m+1}}
  =
  \alpha_0 \ket{0} \ket{\xi_0}
  +
  \alpha_1 \ket{1} \ket{\xi_1}
}$
for some states $\ket{\xi_0}$ and $\ket{\xi_1}$ in ${(\sfV, \sfM, \sfB)}$
orthogonal to each other,
where
${
  \alpha_1
  =
  \sqrt{p_{\acc}} \cdot \sqrt{\frac{1-\varepsilon}{p_{\acc}}}
  =
  \sqrt{1 - \varepsilon}
}$
and
${
  \alpha_0
  =
  \sqrt{1 - \abs{\alpha_1}^2}
  =
  \sqrt{\varepsilon}
}$.

At the ${\left( \frac{m}{2} + 1 \right)}$-st transformation of $R$,
after receiving the registers $\sfV$, $\sfM$, and $\sfB$ from $W$,
$R$ applies the unitary transformation $Z$
to the qubits in ${(\sfV, \sfM, \sfB)}$
such that
${Z \ket{\xi_0} = \ket{\eta} \ket{0}}$
and
${Z \ket{\xi_1} = \ket{\eta} \ket{1}}$
for some state $\ket{\eta}$ in ${(\sfV, \sfM)}$
(this is possible because $\ket{\xi_0}$ and $\ket{\xi_1}$ are orthogonal).
$R$ then sends $\sfB$ to $W$,
which is the last message of the constructed protocol.

Now the perfect completeness is obvious
from the constructions of $W$ and $R$.

Finally, the zero-knowledge property against $W$ is almost straightforward.

Let $S_V$ be the simulator for the original $m$-message system
such that,
if $x$ is in $A_{\yes}$,
the states ${S_V(x,j)}$ and ${\view_{V,P}(x,j)}$
are computationally indistinguishable,
for each ${1 \leq j \leq \frac{m}{2}}$.

The simulator $T_W$ for the constructed ${(m+2)}$-message system
behaves as follows.

Let ${T_W(x,j)}$ be a quantum state in ${(\sfX, \sfV, \sfM)}$
defined by ${T_W(x,j) = \ketbra{0} \otimes S_V(x,j)}$
for each ${1 \leq j \leq \frac{m}{2}-1}$.
Let ${T_W \bigl( x, \frac{m}{2} \bigr)}$ be a quantum state
in ${(\sfX, \sfV, \sfM, \sfB)}$
defined by
${
  T_W \bigl( x, \frac{m}{2} \bigr)
  =
  \ketbra{0} \otimes S_V \bigl( x, \frac{m}{2} \bigr) \otimes \ketbra{1}
}$.
Finally, let ${T_W \bigl( x, \frac{m}{2}+1 \bigr)}$ be a quantum state
in ${(\sfX, \sfB)}$
defined by
${
  T_W \bigl( x, \frac{m}{2}+1 \bigr)
  =
  \ketbra{\phi}
}$.
It is obvious that the ensemble ${\{ T_W(x,j) \}}$
is polynomial-time preparable.

Suppose that $x$ is in $A_{\yes}$.
For ${1 \leq j \leq \frac{m}{2}-1}$,
${T_W(x,j)}$ is obviously computationally indistinguishable from ${\view_{W,R}(x,j)}$,
since
${T_W(x,j) = \ketbra{0} \otimes S_V(x,j)}$,
${\view_{W,R}(x,j) = \ketbra{0} \otimes \view_{V,P}(x,j)}$,
and ${S_V(x,j)}$ and ${\view_{V,P}(x,j)}$
are computationally indistinguishable.
The computational indistinguishability between
${T_W \bigl( x, \frac{m}{2} \bigr)}$
and
${\view_{W,R} \bigl( x, \frac{m}{2} \bigr)}$
follows from
the computational indistinguishability between
${S_V \bigl( x, \frac{m}{2} \bigr)}$
and
${\view_{V,P} \bigl( x, \frac{m}{2} \bigr)}$
and the fact that
${
  \trnorm{
    \view_{W,R} \bigl( x, \frac{m}{2} \bigr)
    -
    \ketbra{0} \otimes \view_{V,P} \bigl( x, \frac{m}{2} \bigr) \otimes \ketbra{1}}
  =
  \trnorm{\ketbra{b} - \ketbra{1}}
  \leq
  2 \sqrt{1 - \frac{1-\varepsilon}{p_{\acc}}}
  \leq
  2 \sqrt{\varepsilon}
}$
is negligible.
Finally,
${T_W \bigl( x, \frac{m}{2}+1 \bigr)}$
and
${\view_{W,R} \bigl( x, \frac{m}{2}+1 \bigr)}$
are identical,
and thus, are trivially computationally indistinguishable.
\end{proof}

Together with Lemmas~\ref{Lemma: three-message public-coin HVQZK systems}~and~\ref{Lemma: computational zero-knowledge against any verifier}
and the computational zero-knowledge version of
Lemma~\ref{Lemma: parallel repetition of three-message HVQPZK},
this implies the equivalence
between quantum computational zero-knowledge with perfect completeness
and usual quantum computational zero-knowledge with two-sided bounded error.
The proof is similar to those of Theorems~\ref{Theorem: HVQPZK = QPZK}~and~\ref{Theorem: HVQZK = QZK}.

\begin{theorem}
Any problem in $\QZK$
has a quantum computational zero-knowledge proof system of perfect completeness.
\label{Theorem: QZK = QZK with perfect completeness}
\end{theorem}

Furthermore, in the computational zero-knowledge case,
it is straightforward
to extend Lemma~\ref{Lemma: computational zero-knowledge against any verifier}
to the following more general statement.

\begin{lemma}
Any three-message public-coin 
honest-verifier quantum computational zero-knowledge system
such that the message from the verifier
consists of ${O(\log n)}$ bits for every input of length $n$
is computational zero-knowledge against
any polynomial-time quantum verifier.
\label{Lemma: computational zero-knowledge against any verifier -- general statement}
\end{lemma}

Using Lemma~\ref{Lemma: computational zero-knowledge against any verifier -- general statement},
we can show the following.

\begin{theorem}
Any problem in $\QZK$
has a three-message public-coin quantum computational zero-knowledge proof system
of perfect completeness
with soundness error probability at most $\frac{1}{p}$
for any polynomially bounded function
$\function{p}{\Nonnegative}{\Natural}$
(hence with arbitrarily small constant error in soundness).
\label{Theorem: three-message public-coin QZK}
\end{theorem}

\begin{proof}
Let $\function{p}{\Nonnegative}{\Natural}$
be any polynomially bounded function,
and let $\function{q}{\Nonnegative}{\Natural}$
be a polynomially bounded function
satisfying ${2^{\frac{q}{2}} \geq \log p + 2}$.

Then,
from Lemmas~\ref{Lemma: HVQZK with perfect completeness}~and~\ref{Lemma: reducing number of messages of HVQZK to three -- modified KW}
together with Lemma~\ref{Lemma: parallel repetition of three-message HVQZK}
for parallel repetition,
we have that
${\HVQZK \subseteq \HVQZK(3, 1, 2^{-q})}$.

With Lemma~\ref{Lemma: three-message public-coin HVQZK systems},
this further implies that
any problem in $\HVQZK$
has a three-message public-coin
honest-verifier quantum computational zero-knowledge proof system
of perfect completeness
with soundness accepting probability at most
${\frac{1}{2} + 2^{-\frac{q}{2}-1}}$
in which the message from the verifier
consists of only one classical bit.

For every input of length $n$,
we run this proof system ${\ceil{\log p(n)} + 2}$ times in parallel.
From Lemma~\ref{Lemma: parallel repetition of three-message HVQZK},
this results in a three-message public-coin
honest-verifier computational zero-knowledge proof system
of perfect completeness
with soundness accepting probability at most
${
  \frac{1}{4p(n)} \bigl( 1+2^{-\frac{q(n)}{2}} \bigr)^{\ceil{\log p(n)} +2}
  \leq
  \frac{1}{p(n)}
}$
in which the message of the verifier
consists of ${\ceil{\log p(n)} + 2}$ classical bits,
for every input of length $n$.

Now Lemma~\ref{Lemma: computational zero-knowledge against any verifier -- general statement}
implies that
this protocol is computational zero-knowledge
even against any dishonest quantum verifier.
Hence, any problem in $\QZK$
has a three-message public-coin quantum computational zero-knowledge proof system
of perfect completeness
with soundness error probability at most $\frac{1}{p}$,
since ${\HVQZK = \QZK}$ by Theorem~\ref{Theorem: HVQZK = QZK}.
\end{proof}


\section{Statistical Zero-Knowledge Case}
\label{Section: QSZK}

All the properties shown for the computational zero-knowledge case
also hold for the statistical zero-knowledge case.
The proofs are essentially same as in the computational zero-knowledge case.
This gives alternative proofs for the following theorems,
which were originally shown by Watrous~\cite{Wat06STOC}
using his previous results~\cite{Wat02FOCS}.

\begin{theorem}[\cite{Wat02FOCS, Wat06STOC}]
${\HVQSZK = \QSZK}$.
\label{Theorem: HVQSZK = QSZK}
\end{theorem}

\begin{theorem}[\cite{Wat02FOCS, Wat06STOC}]
Any problem in $\QSZK$
has a public-coin quantum statistical zero-knowledge proof system.
\label{Theorem: public-coin QSZK = QSZK}
\end{theorem}

We also have the following new properties
for quantum statistical zero-knowledge.

\begin{theorem}
Any problem in $\QSZK$
has a quantum statistical zero-knowledge proof system of perfect completeness.
\label{Theorem: QSZK = QSZK with perfect completeness}
\end{theorem}

\begin{theorem}
Any problem in $\QSZK$
has a three-message public-coin quantum statistical zero-knowledge proof system
of perfect completeness
with soundness error probability at most $\frac{1}{p}$
for any polynomially bounded function
$\function{p}{\Nonnegative}{\Natural}$
(hence with arbitrarily small constant error in soundness).
\label{Theorem: three-message public-coin QSZK}
\end{theorem}


\section{Equivalence of Two Definitions of Quantum Perfect Zero-Knowledge}
\label{Section: Equivalence of Two Definitions of Quantum Perfect Zero-Knowledge}

In the classical case,
the most common definition of perfect zero-knowledge proofs
seems to allow the simulator to output ``$\FAIL$''
with small probability,
say, with probability at most $\frac{1}{2}$~\cite{Gol01Book, SahVad03JACM}.
Following this convention,
we may consider the following alternative definitions of
honest-verifier and general quantum perfect zero-knowledge proof systems.

\begin{definition}
Given a polynomially bounded function
$\function{m}{\Nonnegative}{\Natural}$
and functions $\function{c, s}{\Nonnegative}{[0,1]}$,
a problem ${A = \{ A_{\yes}, A_{\no} \}}$ is in ${\HVQPZK'(m,c,s)}$
iff there exists an $m$-message honest quantum verifier $V$
and an $m$-message honest quantum prover $P$
such that
\begin{description}
\item[\textnormal{(Completeness and Soundness)}]
 ${(V,P)}$ forms an $m$-message quantum interactive proof system
with completeness accepting probability at least $c$
and
soundness accepting probability at most $s$,
\item[\textnormal{(Honest-Verifier Perfect Zero-Knowledge)}]
there exists a polynomial-time preparable ensembles ${\{S_V(x,j)\}}$ of quantum states
such that,
for every ${x \in A_{\yes}}$ and for each ${1 \leq j \leq \bigceil{\frac{m(\abs{x})}{2}}}$,
${
  S_V(x,j)
  =
  p_{x,j} \ketbra{0} \otimes \ketbra{0_{\calH_j}}
  + (1-p_{x,j}) \ketbra{1} \otimes \view_{V,P}(x,j)
}$
for some ${0 \leq p_{x,j} \leq \frac{1}{2}}$,
where $\calH_j$ is the Hilbert space
${\view_{V,P}(x,j)}$ is in ${\Density(\calH_j)}$.
\end{description}
\label{Definition: HVQPZK'(m,c,s)}
\end{definition}

\begin{definition}
Given a polynomially bounded function
$\function{m}{\Nonnegative}{\Natural}$
and functions $\function{c, s}{\Nonnegative}{[0,1]}$,
a problem ${A = \{ A_{\yes}, A_{\no} \}}$ is in ${\QPZK'(m,c,s)}$
iff there exists an $m$-message honest quantum verifier $V$
and an $m$-message honest quantum prover $P$
such that
\begin{description}
\item[\textnormal{(Completeness and Soundness)}]
 ${(V,P)}$ forms an $m$-message quantum interactive proof system
with completeness accepting probability at least $c$
and
soundness accepting probability at most $s$,
\item[\textnormal{(Perfect Zero-Knowledge)}]
for any $m$-message quantum verifier $V'$,
there exists a polynomial-time uniformly generated family ${\{Q_x\}}$ of quantum circuits,
where each $Q_x$ exactly implements an admissible transformation $S_{V'}(x)$,
such that,
for every ${x \in A_{\yes}}$,
${
  S_{V'}(x)
  =
  p_x (\Phi_0 \otimes \Psi_{\fail})
  +
  (1-p_x) (\Phi_1 \otimes \lrangle{V',P}(x))
}$
for some ${0 \leq p_x \leq \frac{1}{2}}$,
where
${\lrangle{V',P}(x) \in \Admissible(\calA, \calZ)}$
is the induced admissible transformation from $V'$, $P$, and $x$
for some Hilbert spaces $\calA$ and $\calZ$,
${\Psi_{\fail} \in \Admissible(\calA, \calZ)}$
is the admissible transformation
that always outputs $\ketbra{0_{\calZ}}$,
and $\Phi_b$
is the admissible transformation
that takes nothing as input and outputs $\ketbra{b}$,
for each ${b \in \Binary}$.
\end{description}
\label{Definition: QPZK'(m,c,s)}
\end{definition}

In Definitions~\ref{Definition: HVQPZK'(m,c,s)}~and~\ref{Definition: QPZK'(m,c,s)},
the first qubit of the output of the simulator
indicates whether or not the simulation succeeds
--- $\ketbra{0}$ is interpreted as failure
and $\ketbra{1}$ as success.

\begin{definition}
A problem ${A = \{ A_{\yes}, A_{\no} \}}$ is in
$\HVQPZK'$ and in $\QPZK'$
if there exists a polynomially bounded function
$\function{m}{\Nonnegative}{\Natural}$
such that $A$ is in
${\HVQPZK' \left(m, \frac{2}{3}, \frac{1}{3} \right)}$
and in ${\QPZK' \left(m, \frac{2}{3}, \frac{1}{3} \right)}$,
respectively.
\label{Definition: HVQPZK' and QPZK'}
\end{definition}

It is not obvious at a glance
that ${\HVQPZK = \HVQPZK'}$ and ${\QPZK = \QPZK'}$,
i.e., that the definitions of
honest-verifier and general quantum perfect zero-knowledge proof systems
using Definitions~\ref{Definition: HVQPZK(m,c,s)}~and~\ref{Definition: QPZK(m,c,s)}
is equivalent to those using Definitions~\ref{Definition: HVQPZK'(m,c,s)}~and~\ref{Definition: QPZK'(m,c,s)}.

Fortunately,
using Theorem~\ref{Theorem: HVQPZK = QPZK},
we can show that ${\HVQPZK = \HVQPZK'}$ and ${\QPZK = \QPZK'}$.
It is stressed that such equivalence is not known in the classical case.

\begin{theorem}
${\HVQPZK = \HVQPZK'}$ and ${\QPZK = \QPZK'}$.
\end{theorem}

\begin{proof}
It is obvious that
${\HVQPZK \subseteq \HVQPZK'}$
and
${\QPZK \subseteq \QPZK' \subseteq \HVQPZK'}$.
From Theorem~\ref{Theorem: HVQPZK = QPZK},
we have ${\HVQPZK = \QPZK}$.
Therefore, it is sufficient to show that ${\HVQPZK' \subseteq \HVQPZK}$.

Let ${A = \{A_{\yes}, A_{\no}\}}$ be a problem
in ${\HVQPZK' \left(m, \frac{2}{3}, \frac{1}{3} \right)}$
for some polynomially bounded function
$\function{m}{\Nonnegative}{\Natural}$.
Without loss of generality,
it is assumed that $m$ takes only even values
(if ${m(n)}$ is odd for some ${n \in \Nonnegative}$,
 we modify the protocol so that the verifier sends
 a ``dummy'' message to a prover as the first message
 when the input has length $n$ such that ${m(n)}$ is odd).
Let $V$ and $P$ be the corresponding honest verifier and honest prover,
respectively.
Let $\sfV$ be the quantum register
consisting of all the qubits in the private space of $V$,
and let $\sfM$ be that
consisting of all the qubits in the message channel between $V$ and the prover.
For every input $x$,
$V$ applies $V_j$ for his $j$th transformation
to the qubits in ${(\sfV, \sfM)}$
for ${1 \leq j \leq \frac{m}{2} + 1}$,
and performs the measurement ${\Pi = \{ \Pi_{\acc}, \Pi_{\rej} \}}$
at the end of the original protocol to decide acceptance or rejection.
Let $\calV$ and $\calM$ be the Hilbert spaces
corresponding to $\sfV$ and $\sfM$, respectively.

Let ${\{S_V(x,j)\}}$ be the polynomial-time preparable ensembles of quantum states
corresponding to the simulator
for this honest-verifier quantum perfect zero-knowledge proof system
such that,
for every ${x \in A_{\yes}}$ and for each ${1 \leq j \leq \frac{m(\abs{x})}{2}}$,
${
  S_V(x,j)
  = p_{x,j} \ketbra{0} \otimes \ketbra{0_{\calV \otimes \calM}}
 + (1-p_{x,j}) \ketbra{1} \otimes \view_{V,P}(x,j)
}$
for some ${0 \leq p_{x,j} \leq \frac{1}{2}}$.
This may be viewed as ${S_V(x,j)}$ outputting
${\ketbra{0} \otimes \ketbra{0_{\calV \otimes \calM}}}$
with probability ${p_{x,j}}$
and ${\ketbra{1} \otimes \view_{V,P}(x,j)}$
with probability ${1 - p_{x,j}}$.
Without loss of generality, it is assumed that
each ${0 \leq p_{x,j} \leq 2^{-\abs{x}}}$,
since we can easily amplify the success probability of the simulator
by just running the original simulator a number of times
so that a new simulator
outputs ${\ketbra{0} \otimes \ketbra{0_{\calV \otimes \calM}}}$
only if all the attempts result in
${\ketbra{0} \otimes \ketbra{0_{\calV \otimes \calM}}}$.

First we slightly modify the behavior of the honest verifier as follows
(call this modified honest verifier $V'$).
At the beginning of the protocol,
$V'$ prepares a single-qubit quantum register $\sfB$
in addition to the registers $\sfV$ and $\sfM$.
The content of $\sfB$ will denote
if the protocol successfully simulates the original protocol
(that $\sfB$ contains $1$
indicates the successful simulation).
At the first transformation of $V'$,
$V'$ prepares $\ket{1}$ in $\sfB$
and ${V_1 \ket{0_{\calV \otimes \calM}}}$ in ${(\sfV, \sfM)}$,
and sends $\sfB$ and $\sfM$ to a prover.
At every message from the prover,
$V'$ receives $\sfB$
in addition to the qubits in $\sfM$ the original verifier $V$ would receive.
At the $j$th transformation of $V'$,
$V'$ applies $V_j$ to the qubits in ${(\sfV, \sfM)}$,
for ${2 \leq j \leq \frac{m(\abs{x})}{2}+1}$.
That is, the $j$th transformation of $V'$
is given by
${
  V'_j
  =
  I \otimes V_j
}$,
for ${2 \leq j \leq \frac{m(\abs{x})}{2}+1}$.
Then $V'$ sends $\sfB$ and $\sfM$ back to the prover
as the ${(2j-1)}$-st message,
for ${2 \leq j \leq \frac{m(\abs{x})}{2}}$.
At the end of the protocol,
$V'$ accepts if and only if
the content of $\sfB$ is $1$
and the content of ${(\sfV, \sfM)}$ corresponds to
an accepting state of the original protocol.

It is obvious that the soundness accepting probability
is at most $\frac{1}{3}$,
since it cannot be larger than that in the original protocol
from the construction of $V'$.

To show the completeness and honest-verifier perfect zero-knowledge conditions,
we construct a new honest prover $P'$ as follows.
Let $\sfP$ be the quantum register
consisting of all the qubits
in the private space of the original honest prover $P$.
The new prover $P'$ prepares $\sfP$
as well as single-qubit quantum registers $\sfB'_j$
and quantum registers $\sfV'_j$ and $\sfM'_j$
in his private space
for ${1 \leq j \leq \frac{m(\abs{x})}{2}}$,
where $\sfV'_j$ and $\sfM'_j$ consists of the same number of qubits
as $\sfV$ and $\sfM$, respectively.
All the qubits in the registers $\sfP$, $\sfB'_j$, $\sfV'_j$, and $\sfM'_j$,
for ${1 \leq j \leq \frac{m(\abs{x})}{2}}$,
are initialized to state $\ket{0}$.

At the $j$th transformation of $P'$,
for ${1 \leq j \leq \frac{m(\abs{x})}{2}}$,
after having received $\sfB$ and $\sfM$,
$P'$ first measures the qubit in $\sfB$
in the ${\{ \ket{0}, \ket{1} \}}$ basis
to obtain the measurement outcome $b$.

If ${b=0}$,
$P'$ does nothing and just sends $\sfB$ and $\sfM$ back to the verifier.

On the other hand, if ${b=1}$,
$P'$ first generates ${S_V(x,j)}$ in ${(\sfB'_j, \sfV'_j, \sfM'_j)}$.
If this results in 
${\ketbra{0} \otimes \ketbra{0_{\calV \otimes \calM}}}$,
$P'$ flips the content of $\sfB$
so that $\sfB$ now contains $0$,
and sends $\sfB$ and $\sfM$ back to the verifier.
Otherwise $P'$ applies $P_j$,
the $j$th transformation of the original honest prover $P$,
to the qubits in ${(\sfM, \sfP)}$,
and sends $\sfB$ and $\sfM$ back to the verifier
(note that $\sfB$ always contains $1$ in this case).

From the construction of $P'$,
it is easy to see that,
if the input $x$ is in $A_{\yes}$,
$P'$ is accepted with probability at least
${\frac{2}{3} (1 - 2^{-\abs{x}})^{\frac{m(\abs{x})}{2}} \geq \frac{5}{9}}$.

Next we construct a new simulator $S'_{V'}$ as follows.
$S'_{V'}$ prepares the quantum registers
$\sfB$, $\sfV$, and $\sfM$
and another three quantum registers $\sfB'$, $\sfV'$, and $\sfM'$,
where $\sfB'$, $\sfV'$, and $\sfM'$ consists of
the same number of qubits as $\sfB$, $\sfV$, and $\sfM$, respectively.
For convenience,
let ${S'_{V'}(x,0) = \ketbra{1} \otimes \ketbra{0_{\calV \otimes \calM}}}$.
We define $S'_{V'}$ inductively with respect to $j$,
for ${1 \leq j \leq \frac{m(\abs{x})}{2}}$.

Assume that the state ${S'_{V'}(x,j-1)}$ has already been defined.
To simulate the state after the $j$th transformation of $P'$,
$S'_{V'}$ first generates
${
  \rho_j
  =
  V'_j S'_{V'}(x,j-1) \conjugate{{V'_j}}
}$
in ${(\sfB, \sfV, \sfM)}$.
If the content of $\sfB$ is $0$,
$S'_{V'}$ just outputs the state in ${(\sfB, \sfV, \sfM)}$.
Otherwise if the content of $\sfB$ is $1$,
$S'_{V'}$ generates the state ${S_V(x,j)}$ in ${(\sfB', \sfV', \sfM')}$.
If the content of $\sfB'$ is $0$,
$S'_{V'}$ outputs the state in ${(\sfB', \sfV, \sfM)}$,
otherwise if the content of $\sfB'$ is $1$,
$S'_{V'}$ outputs the state in ${(\sfB', \sfV', \sfM')}$.

Let $\Pi_b$ be the projection defined by
${\Pi_b = \ketbra{b} \otimes I_{\calV \otimes \calM}}$,
for each ${b \in \Binary}$.
Then, ${S'_{V'}(x,j)}$ can be written as
\[
\begin{split}
S'_{V'}(x,j)
&
=
\Pi_0 \rho_j \Pi_0
+
(\tr \Pi_0 S_V(x,j)) \ketbra{0} \otimes \tr_{\calB} \Pi_1 \rho_j \Pi_1
+
(\tr \Pi_1 \rho_j) \Pi_1 S_V(x,j) \Pi_1
\\
&
=
\Pi_0 \rho_j \Pi_0
+
p_{x,j} \ketbra{0} \otimes \tr_{\calB} \Pi_1 \rho_j \Pi_1
+
(\tr \Pi_1 \rho_j) (1 - p_{x,j}) \ketbra{1} \otimes \view_{V,P}(x,j),
\end{split}
\]
for ${1 \leq j \leq \frac{m(\abs{x})}{2}}$,
where ${\calB}$ is the Hilbert space corresponding to $\sfB$.

It is easy to see that the ensemble ${\{ S'_{V'}(x,j) \}}$
is polynomial-time preparable.

Suppose that $x$ is in $A_{\yes}$.
We show by induction that ${S'_{V'}(x,j) = \view_{V',P'}(x,j)}$
for each ${1 \leq j \leq \frac{m(\abs{x})}{2}}$.
For convenience,
let ${\view_{V',P'}(x,0) = S'_{V'}(x,0) = \ketbra{1} \otimes \ketbra{0_{\calV \otimes \calM}}}$,
and let
${
  \sigma_j
  =
  V'_j \view_{V',P'}(x,j-1) \conjugate{{V'_j}}
}$
for each ${1 \leq j \leq \frac{m(\abs{x})}{2}}$.

In the case ${j=1}$,
it is obvious that ${S'_{V'}(x,1) = \view_{V',P'}(x,1)}$,
since
\[
\rho_1
=
\sigma_1
=
V'_1 (\ketbra{1} \otimes \ketbra{0_{\calV \otimes \calM}}) \conjugate{{V'_1}}
=
\ketbra{1} \otimes (V_1 \ketbra{0_{\calV \otimes \calM}} \conjugate{V_1}),
\]
and thus
\[
\begin{split}
S'_{V'}(x,1)
&=
p_{x,1} \ketbra{0} \otimes \tr_{\calB} \Pi_1 \rho_1 \Pi_1
+
(1 - p_{x,1}) \ketbra{1} \otimes \view_{V,P}(x,1)
\\
&
=
p_{x,1} \ketbra{0} \otimes \tr_{\calB} \Pi_1 \sigma_1 \Pi_1
+
(1 - p_{x,1}) \ketbra{1} \otimes \view_{V,P}(x,1)
\\
&
=
\view_{V',P'}(x,1).
\end{split}
\]

Suppose that ${S'_{V'}(x,j) = \view_{V',P'}(x,j)}$ holds
for all ${1 \leq j \leq k}$.
We show the case ${j = k+1}$.
By definition,
\[
S'_{V'}(x,k+1)
=
\Pi_0 \rho_{k+1} \Pi_0
+
(\tr \Pi_0 S_V(x,k+1)) \ketbra{0} \otimes \tr_{\calB} \Pi_1 \rho_{k+1} \Pi_1
+
(\tr \Pi_1 \rho_{k+1}) \Pi_1 S_V(x,k+1) \Pi_1,
\]
and notice that
\[
\begin{split}
\view_{V',P'}(x,k+1)
&
=
\Pi_0 \sigma_{k+1} \Pi_0
+
(\tr \Pi_0 S_V(x,k+1)) \ketbra{0} \otimes \tr_{\calB} \Pi_1 \sigma_{k+1} \Pi_1
\\
&
\hspace{1cm}
+
(\tr \Pi_1 \sigma_{k+1}) (\tr \Pi_1 S_V(x,k+1)) \ketbra{1} \otimes \view_{V,P}(x,k+1).
\end{split}
\]
Since
${
  \rho_{k+1}
  =
  V'_{k+1} S'_{V'}(x,k) \conjugate{{V'_{k+1 \hspace{-3ex}}}}
  \hspace{1.25ex}
}$
and
${
  \sigma_{k+1}
  =
  V'_{k+1} \view_{V',P'}(x,k) \conjugate{{V'_{k+1 \hspace{-3ex}}}}
  \hspace{1.25ex}
}$,
we have ${\rho_{k+1} = \sigma_{k+1}}$
from the assumption that ${S'_{V'}(x,k) = \view_{V',P'}(x,k)}$.
Furthermore, we have
\[
\Pi_1 S_V(x,k+1) \Pi_1
=
(\tr \Pi_1 S_V(x,k+1)) \ketbra{1} \otimes \view_{V,P}(x,k+1).
\]
Therefore, that ${S'_{V'}(x,k+1) = \view_{V',P'}(x,k+1)}$ follows.

Hence, the honest-verifier perfect zero-knowledge property against $P'$ holds
in the sense of Definition~\ref{Definition: HVQPZK(m,c,s)}.

Finally, recall that the success probability
can be amplified using sequential repetition,
and thus, that ${\HVQPZK' \subseteq \HVQPZK}$ follows.
\end{proof}


\section{Conclusion}
\label{Section: conclusion}

This paper has established
a unified framework that directly proves a number of general properties
of quantum zero-knowledge proofs.
Our method works well
for any of quantum perfect, statistical, and computational
zero-knowledge cases.
We conclude by mentioning several open problems
concerning quantum zero-knowledge proofs:
\begin{itemize}
\item
  We have proved that quantum computational and statistical zero-knowledge proofs
  can be made perfect complete.
  Can quantum \emph{perfect} zero-knowledge proofs
  be made perfect complete?
\item
  Although we have proved properties of quantum zero-knowledge proofs directly,
  natural complete problems or characterizations are definitely helpful
  when proving properties of quantum zero-knowledge proofs.
  Are their any natural complete problems or characterizations
  for $\QZK$ and $\QPZK$?
\item
  We have investigated the properties of $\QZK$ that hold unconditionally.
  On the other hand, Watrous~\cite{Wat06STOC} proved that
  every problem in $\NP$
  has a quantum computational zero-knowledge proof system
  under some intractability assumptions.
  In the classical case,
  it is known that every problem in ${\IP = \PSPACE}$
  is provable in computational zero-knowledge
  under some intractability assumptions~\cite{ImpYun87CRYPTO, BenGolGolHasKilMicRog88CRYPTO, LunForKarNis92JACM, Sha92JACM}.
  How powerful are quantum computational zero-knowledge proofs
  under reasonable intractability assumptions?
\end{itemize}


\section*{Acknowledgement}

The author would like to thank John Watrous
for his helpful comments on the choice of the universal gate set.




\appendix

\section*{\appendixname}


\section{Quantum Interactive Proof Systems}
\label{Appendix: QIP model}

Here we review the model of quantum interactive proof systems.
Although the term ``round'' is commonly used
in classical interactive proofs
for describing each set of verifier's question
and corresponding prover's response,
this paper follows the custom in the preceding papers
of quantum interactive proofs~\cite{Wat03TCS, KitWat00STOC, Wat02FOCS, MarWat05CC, RosWat05CCC}
and uses the term ``message'' instead of ``round''.
One round consists of two messages:
the message from a verifier and the message from a prover.

A quantum interactive proof system consists of two parties:
a quantum verifier $V$ and a quantum prover $P$.
Associated with the quantum interactive proof system
are the Hilbert spaces $\calV$, $\calM$, and $\calP$,
where $\calV$ corresponds to the private space of the verifier $V$,
$\calM$ corresponds to the space used for communication
between the verifier $V$ and the prover $P$,
and $\calP$ corresponds to the private space of the prover $P$.

For every input of length $n$,
each space $\calV$, $\calM$, and $\calP$
consists of ${q_{\calV}(n)}$, ${q_{\calM}(n)}$, and ${q_{\calP}(n)}$ qubits,
respectively,
for some polynomially bounded functions
$\function{q_{\calV}, q_{\calM}}{\Nonnegative}{\Natural}$
and some function $\function{q_{\calP}}{\Nonnegative}{\Natural}$.
Accordingly, the entire system consists of
${q(n) = q_{\calV}(n) + q_{\calM}(n) + q_{\calP}(n)}$
qubits.
Such a system is called
\emph{${(q_{\calV}, q_{\calM}, q_{\calP})}$-space-bounded},
and the associated verifier and prover are called
\emph{${(q_{\calV}, q_{\calM})}$-space-bounded}
and
\emph{${(q_{\calM}, q_{\calP})}$-space-bounded},
respectively.
One of the private qubits of the verifier is designated as the output qubit.

Formally, an \emph{$m$-message} ${(q_{\calV}, q_{\calM})}$-space-bounded
quantum verifier $V$
for quantum interactive proof systems
is a polynomial-time computable mapping of the form
$\function{V}{\Binary^{\ast}}{\Binary^{\ast}}$.
For every $n$ and for every input ${x \in \Binary^{\ast}}$ of length $n$,
$V$ uses at most ${q_{\calV}(n)}$ qubits for his private space
and at most ${q_{\calM}(n)}$ qubits for each communication with a prover.
The string ${V(x)}$ is interpreted as a $\ceil{(m(n)+1)/2}$-tuple
${(V(x)_1, \ldots, V(x)_{\ceil{(m(n)+1)/2}})}$,
with each ${V(x)_j}$ a description of
a polynomial-time uniformly generated quantum circuit
acting on ${q_{\calV}(n) + q_{\calM}(n)}$ qubits.

Similarly, an \emph{$m$-message} ${(q_{\calM}, q_{\calP})}$-space-bounded
quantum verifier $P$
is a mapping of the form
$\function{P}{\Binary^{\ast}}{\Binary^{\ast}}$.
For every $n$ and for every input ${x \in \Binary^{\ast}}$ of length $n$,
$P$ uses at most ${q_{\calP}(n)}$ qubits for his private space
and at most ${q_{\calM}(n)}$ qubits for each communication with a verifier.
The string ${P(x)}$ is interpreted as a $\ceil{m(n)/2}$-tuple
${(P(x)_1, \ldots, P(x)_{\ceil{m(n)/2}})}$,
with each ${P(x)_j}$ a description of a quantum circuit
acting on ${q_{\calM}(n) + q_{\calP}(n)}$ qubits.
No restrictions are placed on the complexity of the mapping $P$
(i.e., each ${P(x)_j}$ can be an arbitrary unitary transformation).

Given an $m$-message ${(q_{\calV}, q_{\calM})}$-space-bounded quantum verifier $V$,
an $m$-message ${(q_{\calM}, q_{\calP})}$-space-bounded quantum prover $P$,
and an input $x$ of length $n$,
we define a circuit ${(V(x), P(x))}$ acting over
${\calV \otimes \calM \otimes \calP}$
of ${q(n)}$ qubits as follows.
If ${m(n)}$ is odd, circuits
${P(x)_1, V(x)_1, \ldots, P(x)_{(m(n)+1)/2}, V(x)_{(m(n)+1)/2}}$
are applied in sequence,
each ${V(x)_j}$ to ${\calV \otimes \calM}$
and each ${P(x)_j}$ to ${\calM \otimes \calP}$.
If ${m(n)}$ is even, circuits
${V(x)_1, P(x)_1, \ldots, V(x)_{m(n)/2}, P(x)_{m(n)/2}, V(x)_{m(n)/2 + 1}}$
are applied in sequence.

At any given instant, the state of the entire system is
a unit vector in the space ${\calV \otimes \calM \otimes \calP}$.
At the beginning of the protocol,
the system is in the initial state
such that all the qubits in
${\calV \otimes \calM \otimes \calP}$
are in state $\ket{0}$.
In case $V$ and/or $P$ have some auxiliary quantum states
$\rho$ and/or $\sigma$ at the beginning of protocol,
the qubits in the private space of $V$ and/or $P$
corresponding to these auxiliary quantum states
are initialized to $\rho$ and/or $\sigma$, respectively.
In such a case, the state of the entire system
may be in a mixed state in ${\Density(\calV \otimes \calM \otimes \calP)}$,
and the descriptions below are interpreted
in the context of mixed states with proper modifications.

For every input $x$ of length $n$,
the probability ${p_{\acc}(x, V, P)}$ that ${(V, P)}$ accepts $x$
is defined to be the probability
that an observation of the output qubit
in the $\{ \ket{0}, \ket{1} \}$ basis yields $\ket{1}$,
after the circuit ${(V(x), P(x))}$ is applied to
the initial state
${\ket{\psi_{\init}} \in \calV \otimes \calM \otimes \calP}$.
Let $\Pi_{\acc}$ be the projection
onto the space consisting of states
whose output qubit is in state $\ket{1}$.
Then,
${
p_{\acc}(x, V, P)
=
\norm{
  \Pi_{\acc}
  V(x)_{(m(n)+1)/2} P(x)_{(m(n)+1)/2}
  \cdots
  V(x)_1 P(x)_1
  \ket{\psi_{\init}}
}^2}$
if ${m(n)}$ is odd,
and
${
p_{\acc}(x, V, P)
=
\norm{
  \Pi_{\acc}
  V(x)_{m(n)/2 + 1} P(x)_{m(n)/2} V(x)_{m(n)/2}
  \cdots
  P(x)_1 V(x)_1
  \ket{\psi_{\init}}
}^2}$
if ${m(n)}$ is even.

The class of problems having
an $m$-message quantum interactive proof system
with completeness accepting probability at least $c$
and soundness accepting probability at most $s$
is denoted by ${\QIP(m, c, s)}$.
The following is the formal definition of the class ${\QIP(m, c, s)}$.

\begin{definition}
Given a polynomially bounded function
$\function{m}{\Nonnegative}{\Natural}$
and functions $\function{c, s}{\Nonnegative}{[0,1]}$,
a problem ${A = \{ A_{\yes}, A_{\no} \}}$ is in ${\QIP(m, c, s)}$
iff there exist polynomially bounded functions
$\function{q_{\calV}, q_{\calM}}{\Nonnegative}{\Natural}$
and an $m$-message ${(q_{\calV}, q_{\calM})}$-space-bounded quantum verifier $V$
for quantum interactive proof systems
such that, for every $n$ and for every input $x$ of length $n$,
\begin{description}
\item[\textnormal{(Completeness)}]
if ${x \in A_{\yes}}$,
there exist a function
$\function{q_{\calP}}{\Nonnegative}{\Natural}$,
and an $m$-message ${(q_{\calM}, q_{\calP})}$-space-bounded quantum prover $P$
such that ${(V, P)}$ accepts $x$ with probability at least ${c(n)}$,
\item[\textnormal{(Soundness)}]
if ${x \in A_{\no}}$,
for any function
$\function{q'_{\calP}}{\Nonnegative}{\Natural}$,
and any $m$-message ${(q_{\calM}, q'_{\calP})}$-space-bounded quantum prover $P'$,
${(V, P')}$ accepts $x$ with probability at most ${s(n)}$.
\end{description}
\label{Definition: QIP(m,c,s)}
\end{definition}

Next, we introduce the notions of \emph{public-coin} quantum verifiers
and \emph{public-coin} quantum interactive proof systems.
Intuitively,
a quantum verifier for quantum interactive proof systems
is public-coin
if every message from $V$ consists of a sequence of outcomes
of a fair classical coin-flipping.

Formally,
an $m$-message ${(q_{\calV}, q_{\calM})}$-space-bounded quantum verifier $V$
for quantum interactive proof systems
is \emph{public-coin}
if $V$ has the following properties
for every $n$ and for every input $x$ of length $n$.
At the $j$th transformation of $V$ for ${1 \leq j \leq \floor{m(n)/2}}$,
$V$ first receives at most ${q_{\calM}(n)}$ qubits from a prover,
then flips a fair classical coin at most ${q_{\calM}(n)}$ times
to generate a random string $r_j$ of length at most ${q_{\calM}(n)}$,
and sends $r_j$ to the prover.

An $m$-message ${(q_{\calV}, q_{\calM}, q_{\calP})}$-space-bounded
quantum interactive proof system
is \emph{public-coin}
if the associated $m$-message ${(q_{\calV}, q_{\calM})}$-space-bounded
quantum verifier is public-coin.

The class of problems having
an $m$-message public-coin quantum interactive proof system
with completeness accepting probability at least $c$
and soundness accepting probability at most $s$
is denoted by ${\QAM(m, c, s)}$.
The following is the formal definition of the class ${\QAM(m, c, s)}$.

\begin{definition}
Given a polynomially bounded function
$\function{m}{\Nonnegative}{\Natural}$
and functions $\function{c, s}{\Nonnegative}{[0,1]}$,
a problem ${A = \{ A_{\yes}, A_{\no} \}}$ is in ${\QAM(m, c, s)}$
iff there exist polynomially bounded functions
$\function{q_{\calV}, q_{\calM}}{\Nonnegative}{\Natural}$
iff there exist polynomially bounded functions
$\function{q_{\calV}, q_{\calM}}{\Nonnegative}{\Natural}$
and an $m$-message ${(q_{\calV}, q_{\calM})}$-space-bounded
public-coin quantum verifier $V$
for quantum interactive proof systems
such that, for every $n$ and for every input $x$ of length $n$,
\begin{description}
\item[\textnormal{(Completeness)}]
if ${x \in A_{\yes}}$,
there exist a function
$\function{q_{\calP}}{\Nonnegative}{\Natural}$,
and an $m$-message ${(q_{\calM}, q_{\calP})}$-space-bounded quantum prover $P$
such that ${(V, P)}$ accepts $x$ with probability at least ${c(n)}$,
\item[\textnormal{(Soundness)}]
if ${x \in A_{\no}}$,
for any function
$\function{q'_{\calP}}{\Nonnegative}{\Natural}$,
and any $m$-message ${(q_{\calM}, q'_{\calP})}$-space-bounded quantum prover $P'$,
${(V, P')}$ accepts $x$ with probability at most ${s(n)}$.
\end{description}
\label{Definition: QAM(m,c,s)}
\end{definition}


\section{Note on the Choice of Universal Gate Set}
\label{Appendix: Note on the Choice of Universal Gate Set}

When proving statements concerning quantum perfect zero-knowledge proofs
or proofs having perfect completeness,
we assume that our universal gate set satisfies some conditions,
since these ``perfect'' properties may not hold
with an arbitrary universal gate set.

For instance, in the case of
the paper by Kitaev~and~Watrous~\cite{KitWat00STOC},
when we try to implement their parallelization protocol to three messages
by \emph{unitary} quantum circuits,
we need to implement the controlled-unitary operation
controlled by the message index $r$
chosen by the verifier at his first transformation.
If this implementation is not exact,
we may lose the perfect completeness property after the parallelization,
which affects their final statement that any problem in $\QIP$
has a three-message quantum interactive proof system
of \emph{perfect completeness} with exponentially small error in soundness.

Furthermore,
in the case of the paper by Marriott~and~Watrous~\cite{MarWat05CC},
their method of converting any three-message quantum interactive proof system
to a three-message public-coin one
works well only if the original three-message protocol
is implemented with \emph{unitary} quantum circuits.
Thus, their result inherits the problem of how to implement
with unitary circuits
the parallelization protocol due to Kitaev and Watrous~\cite{KitWat00STOC},
when claiming their statement in a final form
that any problem in $\QIP$
has a three-message public-coin quantum interactive proof system
of \emph{perfect completeness} with exponentially small error in soundness
(i.e., ${\QIP \subseteq \QMAM(1, 2^{-p})}$
for any polynomially bounded function $p$).

This is also the case for the present paper,
since we are using both a modified version
of the parallelization protocol due to Kitaev and Watrous~\cite{KitWat00STOC}
and a public-coin technique due to Marriott~and~Watrous~\cite{MarWat05CC}.
In our case,
if the implementations of the controlled-unitary transformations
are not exact,
we may lose the perfect zero-knowledge property after the parallelization,
since the implementations used for the simulator
may differ from those used for the honest verifier.

One direct solution to avoid these problems
is to use such a universal gate set that
(i) the Hadamard and Toffoli gates are exactly implementable
with a constant number of gates in the universal gate set,
and
(ii) given a circuit $Q$ consisting of gates in the universal gate set
that exactly implements a unitary transformation $U$,
we can construct another circuit $Q'$ consisting of gates in the same universal gate set
that exactly implements the controlled-$U$ transformation
such that the size of $Q'$ is bounded by polynomial with respect to
the size of $Q$.
For instance,
if the Toffoli gate is in our universal gate set $\mathfrak{U}$
and the controlled-$U$ gate is necessarily included in $\mathfrak{U}$
for any gate $U$ in $\mathfrak{U}$ not of controlled-unitary type,
the condition (ii) is satisfied.
This is because the controlled-controlled-$U$ operator
is easily realized by the controlled-$U$ and Toffoli gates.
From these observations,
one can see that,
for example,
the set consisting of
the Hadamard gate, the controlled-Hadamard gate, and the Toffoli gate
satisfies both (i) and (ii).

Watrous~\cite{Wat07private} pointed out
that the condition (ii) is actually not necessary for our purpose.
In fact, what we need is a unitary implementation of
the parallelization protocol that does not lose the ``perfect'' properties.
The essence of the Kitaev-Watrous parallelization method
lies in the use of the controlled-swap test.
Note that, if we may assume the condition (i),
the controlled-swap transformation can be implemented exactly.
Now, instead of implementing the controlled-unitary operation
controlled by the message index $r$,
we may implement the following that is sufficient for our purpose.
For simplicity, it is assumed that $r$ is chosen from
the set ${\{0, \ldots, 2^l-1\}}$ for some positive integer $l$
(such an assumption does not lose generality
because we can appropriately add ``dummy'' messages to the underlying protocol
so that the number of messages becomes $2^{l+1}$ in the underlying protocol),
and the unitary transformation $U_r$ is applied when $r$ is chosen.
Suppose $U_r$ acts over $q$ qubits in a register $\sfT$,
for each $r$.
We prepare ancillae of $q$ qubits in a register $\sfA_r$ for each $r$,
and set the control qubits in a register $\sfC$ to the state
${
  \frac{1}{\sqrt{2^l}} \sum_{r=0}^{2^l-1} \ket{r}
}$.
We first swap the content of $\sfT$ and that of $\sfA_r$
when the content of $\sfC$ is $r$, for each $r$
(this can be realized using controlled-swap transformations).
Next we apply ${U_0 \otimes \cdots \otimes U_{2^l-1}}$
to the qubits in ${(\sfA_0, \ldots, \sfA_{2^l-1})}$,
and then we again swap the content of $\sfT$ and that of $\sfA_r$
when the content of $\sfC$ is $r$, for each $r$.
This results in applying
${
  U_0 \otimes \cdots \otimes U_{r-1}
  \otimes I_{2^q}
  \otimes U_{r+1} \otimes \cdots \otimes U_{2^l-1}
}$
to some meaningless quantum state
when the content of $\sfC$ is $r$,
and thus, would not keep the coherence of the quantum state in $\sfC$.
However, recall that
the control part in the Kitaev-Watrous parallelization protocol
is the message index $r$,
which is originally chosen at random \emph{classically}
when we describe the protocol in a non-unitary manner.
Hence such decoherence does not affect the protocol at all,
and we can have the unitary implementation of the protocol
only using the circuits for $U_r$'s and for the controlled-swap operation.
We may also use a similar technique when constructing a simulator.
To avoid unnecessary complication,
now the honest verifier sends all the ancilla qubits in the registers
${\sfA_0, \ldots, \sfA_{2^l-1}}$
to a prover at the second message
in addition to the actual message prescribed in the protocol.
The honest prover just ignores these ancilla qubits
when sending the third message,
and the simulator does not need to simulate the ancilla qubits.
Therefore, all the ``perfect'' properties
claimed in this paper (and ones in Refs.~\cite{KitWat00STOC, MarWat05CC}) hold
with any gate set such that
the Hadamard transformation and any classical reversible transformations
are exactly implementable.
Fortunately, most of the standard gate sets satisfy this condition.
A typical example is the Shor basis~\cite{Sho96FOCS} consisting of
the Hadamard gate, the controlled-$i$-phase-shift gate, and the Toffoli gate.
\end{document}